\documentclass[%
reprint,
superscriptaddress,
 amsmath,amssymb,
 aps,
 prx,
longbibliography,
floatfix,
]{revtex4-2}

\usepackage{graphicx}
\usepackage{dcolumn}
\usepackage{bm}
\usepackage{comment}
\usepackage{xcolor}
\usepackage{braket}
\usepackage{soul}
\usepackage{hyperref}
\usepackage{multirow}
\newcommand{\pe}{p_\mathrm{e}}
\newcommand{\re}{R_\mathrm{e}}

\usepackage{tabularx}
\usepackage{booktabs}
\hypersetup{
    colorlinks=true,
    linkcolor=blue,
    urlcolor=blue,  
    citecolor=blue,
}

\begin{document}

\preprint{APS/123-QED}

\title{Surface Code with Imperfect Erasure Checks}
\author{Kathleen Chang}
\affiliation{Yale University, Department of Physics, New Haven, CT 06520, USA}
\affiliation{Yale Quantum Institute, Yale University, New Haven, Connecticut 06511, USA}
\author{Shraddha Singh}
\affiliation{Yale University, Department of Applied Physics, New Haven, CT 06520, USA}
\affiliation{Yale Quantum Institute, Yale University, New Haven, Connecticut 06511, USA}
\author{Jahan Claes}
\affiliation{Yale University, Department of Applied Physics, New Haven, CT 06520, USA}
\affiliation{Yale Quantum Institute, Yale University, New Haven, Connecticut 06511, USA}
\author{Kaavya Sahay}
\affiliation{Yale University, Department of Applied Physics, New Haven, CT 06520, USA}
\affiliation{Yale Quantum Institute, Yale University, New Haven, Connecticut 06511, USA}
\author{James Teoh}
\affiliation{Quantum Circuits, Inc., 25 Science Park, New Haven, CT 06511, USA}
\author{Shruti Puri}
\affiliation{Yale University, Department of Applied Physics, New Haven, CT 06520, USA}
\affiliation{Yale Quantum Institute, Yale University, New Haven, Connecticut 06511, USA}

\begin{abstract}

Recently, a lot of effort has been devoted towards designing erasure qubits in which dominant physical noise excites leakage states whose population can be detected and returned to the qubit subspace. Interest in these erasure qubits has been driven by studies showing that the requirements for fault-tolerant quantum error correction are significantly relaxed when noise in every gate operation is dominated by erasures. However, these studies assume perfectly accurate erasure checks after every gate operation which generally come with undesirable time and hardware overhead costs. In this work, we investigate the consequences of using an imperfect but overhead-efficient erasure check for fault-tolerant quantum error correction with the surface code. We show that, under physically reasonable assumptions on the imperfect erasure checks, the threshold error rate is still at least over twice that for Pauli noise. We also study the impact of imperfect erasure checks on the effective error distance and find that it degrades the effective distance under a general error model in which a qubit suffers from depolarizing noise when interacting with a leaked qubit. We then identify a more restrictive but realistic noise model for a qubit that interacts with a leaked qubit, under which the effective error distance is twice that for Pauli noise. We apply our analysis to recently proposed superconducting dual-rail erasure qubits and show that achieving good performance surface code quantum memories with relaxed system requirements is possible. 

\end{abstract}

\maketitle

\section{\label{sec:introduction}Introduction}

Fault-tolerant quantum-error correcting codes promise scalable quantum computation as long as the underlying hardware is not too noisy. 
The amount of error suppression with a given code depends on the nature of noise in the underlying hardware~\cite{alferis2008faulttolerant, aliferis2009, brooks2013, stephens2013, webster2015reducing, tuckett2019tailoring, tuckett2018ultra, tuckett2020faulttolerant, bonilla2021xzzx, dua2024clifford, higgott2023improved, xu2023tailored, singh2022highfidelity, darmawan2021practical,gottesman1997stabilizer,grassl1997codes, wu2022erasure, sahay2023high, kubica2023erasure}. For example, consider a {\it Pauli noise channel} and an {\it erasure channel}.
Under Pauli noise, a qubit may undergo a bit-flip or phase-flip unbeknownst to the user. In contrast, under erasures, the user receives a flag when a specific qubit suffers an error. It has been shown that quantum error correction (QEC) protocols in which every erroneous gate is flagged have larger effective distance and threshold compared to when the gates are subject to Pauli noise~\cite{gottesman1997stabilizer,grassl1997codes, wu2022erasure, sahay2023high, kubica2023erasure}. Consequently, theoretical and experimental efforts are being directed towards designing {\it erasure qubits} in which the dominant hardware noise in the system is converted into erasures~\cite{teoh2023dual, kubica2023erasure, wu2022erasure, sahay2023high, PhysRevA.102.022426, PRXQuantum.4.020358, Ma_2023,koottandavida2023erasure, levine2023demonstrating, chou2024demonstrating, degraaf2024midcircuit}.

In the leading proposals for erasure qubits, quantum operations are designed so that the dominant noise takes the qubit to states out of the computational subspace or {\it leakage states}. The population in these states is detected via its effect on the state of an auxiliary mode. This detection of the leaked population, referred to as an {\it erasure check}, heralds the erroneous qubit, converting the dominant noise to erasures. A particularly interesting solid-state erasure qubit platform is the recently proposed superconducting dual-rail (DR) qubit, which has the advantage of relatively fast microwave controls with the existing circuit-QED toolbox~\cite{kubica2023erasure,teoh2023dual,levine2023demonstrating,koottandavida2023erasure, degraaf2024midcircuit,chou2024demonstrating}. While our results are more generally applicable, we tailor our investigation to regimes relevant to the DR qubit in this work. In this platform, the qubit is encoded in the single-excitation manifold of two microwave cavities or two transmons. Additional transmons are required to implement erasure checks. In the case of cavity-based dual-rails, the transmons are also used to implement two-qubit gate operations~\cite{teoh2023dual,tsunoda2023errordetectable}. 

\begin{table*}[ht]
\centering
\begin{tabular}[c]{p{1.8cm} p{1.7cm} p{1.0cm} p{2.0cm} p{2cm} p{4.0cm} p{0.2cm} p{4.0cm} }
\toprule[1pt]
\multicolumn{8}{c}{Temporal resolution $\eta=0.986$}
\\
\toprule[0.5pt]
\\
Spatial resolution & Leakage-induced Pauli & $R_\mathrm{e}$& Threshold \newline[\%] & $d_\mathrm{eff}$ ($3\times 3$) & \multicolumn{3}{c}{Dual-rail Architecture \& No. of transmons ($3\times 3$)}
\\
& & & & &  \hfil Transmon-based & & \hfil Cavity-based
\\
\toprule[1pt]
\\
imperfect & general & $0.98$\newline $0.99$ & $4.16\pm0.03$ \newline $4.49\pm0.02$ & $1.99\pm0.02$ \newline$2.03\pm0.03$ & & &\\
 & & & & & & &\\
perfect & general & $0.98$ & $4.17\pm0.03$ & $2.00\pm0.02$ & & & \\
& & $0.99$ & $4.55\pm0.02$ & $2.00\pm0.02$ &  & \\
 & & & & & & &\\
imperfect & tailored & $0.98$\newline $0.99$ & $4.23\pm0.03$ \newline $4.47\pm0.03$& $2.38\pm0.03$ \newline $2.60\pm0.05$& & & 49 transmons \& 8 readout lines for QND checks built-into gates~\cite{teoh2023dual, tsunoda2023errordetectable}\\
 & & & & & & &\\
perfect & tailored & $0.98$ \newline $0.99$ & $5.47\pm0.03$ \newline $6.01\pm0.03$ & $2.36\pm0.04$ \newline $2.54\pm0.02$& 75 (58) transmons if three (two) transmons per dual-rail \& 17 readout lines for explicit QND checks with false negatives~\cite{kubica2023erasure,levine2023demonstrating} & & 58 transmons \& 17 readout lines for explicit QND checks with false negatives~\cite{koottandavida2023erasure,teoh2023dual,degraaf2024midcircuit} \\
\\
\bottomrule[1pt]
\end{tabular}
\caption{We evaluate thresholds and effective error distances for four models of leakage-induced Pauli errors and imperfect erasure checks. Details of these noise models are described in Section~\ref{sec: Noise Model and Summary of Main Results}. We estimate these quantities at experimentally-motivated parameters for the erasure fraction $\re$, and temporal resolution $\eta$. The quantity $\eta$ quantifies the fraction of two-qubit gate erasures which are detected immediately after the gate before they propagate to other qubits via subsequent gate operations. Recent superconducting dual-rail experiments have observed $\re$ ranging from $0.96$ to $0.98$ and $\eta$ from $0.95$ to $0.99$~\cite{levine2023demonstrating,koottandavida2023erasure}. In the last two columns, we connect these erasure check models to existing superconducting dual-rail architectures. To provide a sense of the hardware costs, we quote the estimated number of transmon-like elements and read-out lines required to build a $3\times3$ rotated code. This transmon count includes data qubits, ancilla qubits, and their tunable couplers (see Appendix~\ref{app: Accounting no of transmons}). The low transmon-element count for cavity-based built-in erasure checks is because, in principle, it only requires one extra transmon per ancilla dual-rail qubit to carry out gates and erasure checks.}
\label{table}
\end{table*}

Unfortunately, in practice, the benefits of erasure qubits are limited due to the time cost of erasure checks. For example, dual-rail erasure checks are expected to take as long to implement as two-qubit gates. Thus, by checking for erasures after every noisy gate the time cost of one error correction cycle will, at the very least, double, which is undesirable. Furthermore, new erasures are introduced during the time of these erasure checks, increasing the error rate per surface code cycle, thereby reducing the efficacy of the code.

An important benefit of erasure qubits emerges from the fact that {\it the effective error distance}, $d_{\mathrm{eff}}$, of codes implemented with qubits suffering from independently and identically distributed (i.i.d) erasure errors is double that than for qubits that suffer from i.i.d. Pauli noise. Here, $d_{\mathrm{eff}}$ is the sub-threshold logical error rate scaling with physical error rate $p$, and the doubling of $d_{\mathrm{eff}}$ results in a substantial reduction in the number of erasure qubits required to reach a target logical error rate compared to qubits with Pauli noise. However, in practice, this benefit may also be limited by the hardware cost of erasure qubits. For example, consider the transmon-based dual-rail erasure qubit in which each qubit is composed of two transmons and requires an additional transmon for erasure checks and contrast it with the conventional transmon qubit, which suffers from Pauli noise. In this case, $d_{\mathrm{eff}}$ for a code increases only by $15\%$ when it is implemented using dual-rail transmons compared to when it is implemented by conventional transmons. When also considering the number of transmons used to implement tunable couplers, there is only a $55\%$ increase in $d_{\mathrm{eff}}$, compared to the naive expectation of a $100\%$ increase one would expect by introducing erasures (see Appendix~\ref{app: Accounting no of transmons}). Consequently, the overhead advantage in terms is dramatically reduced when using them as dual-rail erasure qubits. 

From the above discussion, we see that it is beneficial to reduce the complexity of erasure checks for dual-rail qubits for practical advantage, for example, by designing faster erasure checks or using the recently proposed two-qubit gate operations with built-in erasure checks for cavity dual-rails~\cite{teoh2023dual, tsunoda2023errordetectable}. The latter approach is attractive because it uses fewer erasure checks and thus fewer transmons, and does not significantly increase the code cycle length, thereby minimizing the accumulation of noise.  However, these alternative strategies come at the cost of a decrease in the quality of erasure checks. For example, the false negative rates may increase or the erasure spatial resolution may decrease i.e. the exact location of the leaked qubit becomes unknown. Consequently, in this work, we study the effect of these two imperfections in the erasure checks on the performance of the surface code when used as a quantum memory. We find that, under reasonable assumptions about noise in the superconducting dual-rail platform, it is still possible to achieve QEC advantage with imperfect erasure checks. 

Our paper is organized as follows: Section~\ref{sec: Noise Model and Summary of Main Results}, we introduce the noise model for imperfect erasure detection. For different erasure detection accuracies and noise models, we quote thresholds and effective error distances for physically relevant parameters in superconducting erasure platforms in
Table~\ref{table} and summarize our main results. In Section~\ref{sec: Noise and decoding when erasure detection is imperfect}, we describe how the decoder is calibrated for checks with imperfect spatial and temporal resolution. Our main results are presented in Section~\ref{sec: effective error distance}, which examines the impact of imperfect erasure checks on the effective error distance, and Section~\ref{sec: Impact on Threshold}, which discusses the impact on the threshold. Note that the numerical analysis in this work is performed with unrotated surface codes. However, the threshold results should also be valid for the rotated codes, and we also present theoretical analysis for the logical error rate scaling of the rotated code (see Appendix~\ref{app:rotated codes}).

\section{Imperfect Erasure Detection Framework and Summary of Main Results}
\label{sec: Noise Model and Summary of Main Results}

\begin{figure}
    \centering
    \includegraphics[width=\columnwidth]{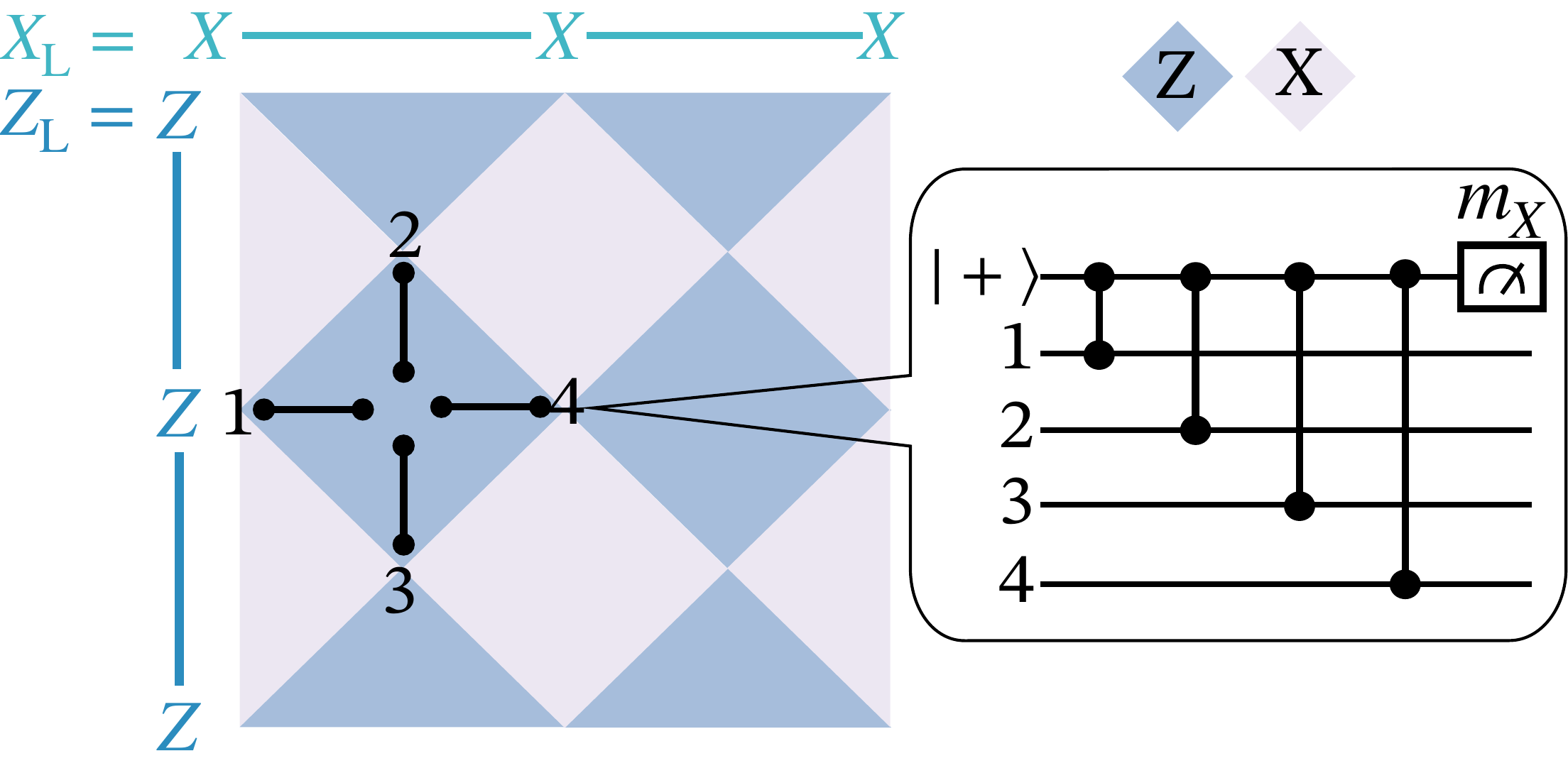}
    \caption{The CSS surface code. Dark (light) plaquettes are  Z (X) stabilizers with support on data qubits at the vertices. The stabilizer is measured by acting 2 qubit gates on a $\ket{+}$-initialized ancilla (not shown) with neighboring data qubits $1-4$ in order. This order is shown for a representative Z stabilizer and is tiled on every stabilizer.}
    \label{fig:noise_models}
\end{figure}

In this section, we introduce the imperfections we consider in erasure checks. 
We consider four models of imperfect erasure checks listed in Table~\ref{table} in a $d\times d$ CSS surface code as depicted in Figure~\ref{fig:noise_models}.  Consider the standard surface code cycle~\cite{DennisTopoQuantumMemory} in which each stabilizer is measured by interacting an ancilla initialized in $\ket{+}$ with the data qubits supported by the stabilizer through CX or CZ gates following a specific order as shown in Figure~\ref{fig:noise_models}(a). Subsequently, the ancilla qubit is measured on the X basis. To simplify the analysis, we only consider erroneous two qubit gates. We do not consider state preparation, measurement, and idling errors because, in principle, they can be combined into the two-qubit gate errors. Given probability $p$ of an error (leakage and Pauli) at a two-qubit gate, we set the probability of leakage during a two-qubit gate as $\pe=\re p$ and the mutually exclusive probability of a Pauli error as $p_\mathrm{p}=(1-R_\mathrm{e}) p$. Here $0\leq \re\leq 1$ is the fraction of leakage errors, or equivalently, $1-\re$ is the fraction of Pauli errors in the gate. The Pauli error, drawn uniformly at random from $\mathcal{E}_\mathrm{p}=\{I,X,Y,Z\}^{\otimes 2}/\{I\otimes I\}$ with probability $p_\mathrm{p}/15$, is applied after a two-qubit gate in simulations. We assume that after every two-qubit gate, an erasure check provides a \textit{flag} to the user indicating if a qubit has leaked. Importantly, we assume that the flag itself is imperfect, which we will describe shortly. This model of (imperfectly) detectable leakage is different from prior studies in which leakage cannot be directly detected and is kept in check with leakage-reduction units or is periodically brought back to the computational subspace and converted to Pauli errors~\cite{10.5555/2011706.2011715,fowler2013coping,ghosh2015leakage,suchara2015leakage,brown2019leakage,battistel2021hardware,mcewen2021removing,marques2023all} 

Before we define these imperfect flags, we first consider what happens when a leaked qubit interacts with another un-leaked qubit before it is detected and reset. We assume a leaked qubit can at most induce a Pauli error on the qubit that it is interacting with~\cite{fowler2013coping,ghosh2013understanding,suchara2015leakage,brown2018comparing,brown2019leakage}. The case in which this leakage-induced Pauli is drawn from $\mathcal{E}_\mathrm{gen.}=\{I,X,Y,Z\}$ with equal probability will be referred to as the {\it general Pauli model}.
We will also analyze a {\it tailored Pauli model} in which the Pauli error induced by the leaked control qubit on the target qubit in a CZ (CX) gate is drawn from $\mathcal{E}_\mathrm{tail.}=\{I,Z\}$ ($\mathcal{E}_\mathrm{tail.}=\{I,X\}$) with equal probability. In the tailored Pauli model, when the target qubit is leaked, the leakage-induced Pauli error on the control qubit is a Pauli drawn from $\{I,Z\}$ with equal probability irrespective of whether a CZ or CX gate was applied. 

The tailored Pauli model was previously studied in the context of leakage reduction units for trapped ions~\cite{brown2019leakage}. Further,
all of the dual-rail gates proposed in~\cite{teoh2023dual}, including the gate with built-in erasure checks~\cite{tsunoda2023errordetectable}, are described by the tailored Pauli model. This is because in the CZ gate, if either one of the qubits undergoes a leakage error, the other qubit may undergo a $Z$-rotation. This noise channel reduces to a dephasing channel under Pauli twirl approximation (PTA) achieved in practice by inserting random single-qubit Pauli gates before and after the CZ gate. The CX gate is implemented by conjugating the target of the CZ gate with Hadamard gates, and hence, its leakage-induced noise channel on the target is the bit-flip channel.  

\begin{figure}
    \centering
    \includegraphics[width=\columnwidth]{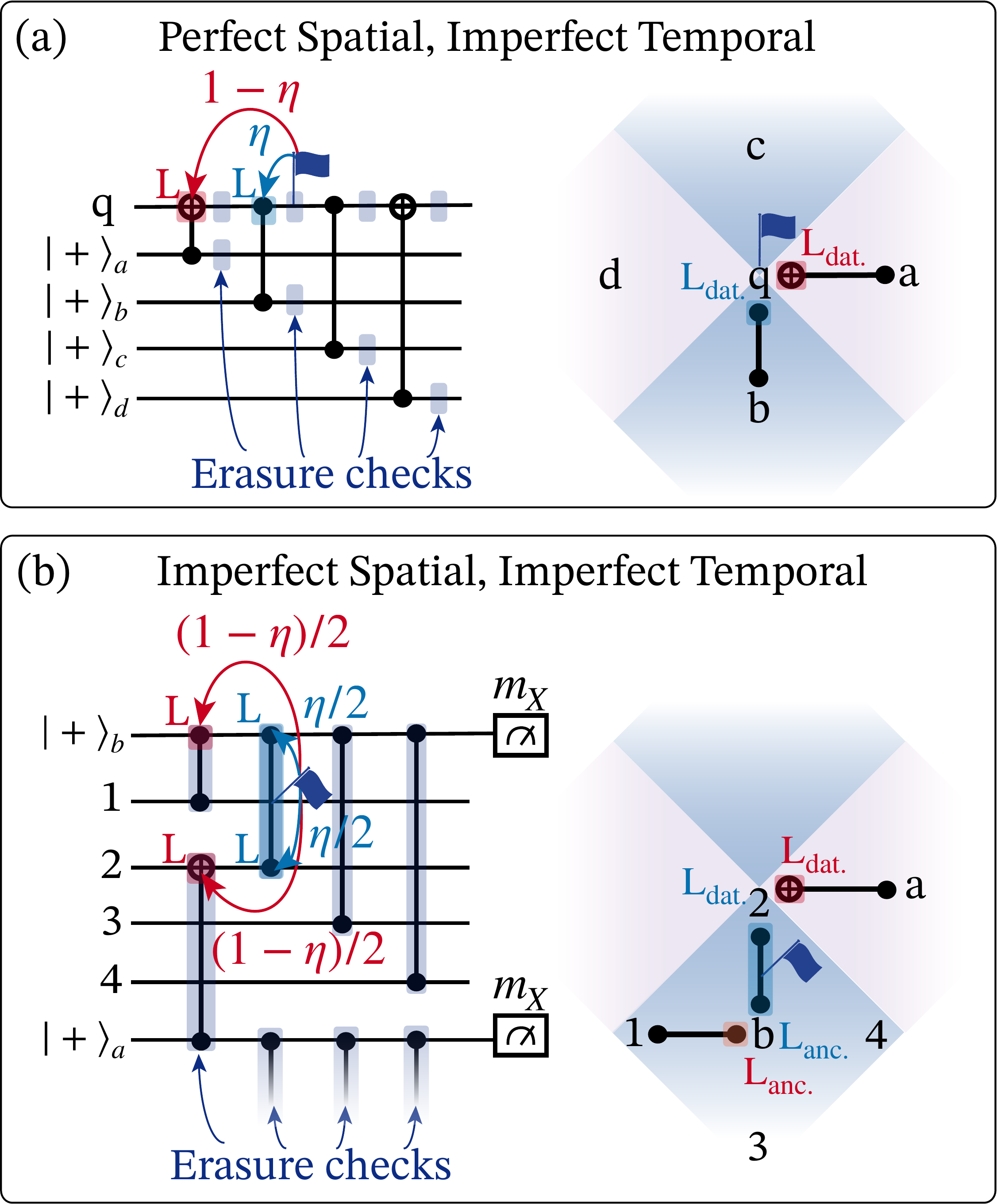}
    \caption{Given checks with (a) perfect spatial and imperfect temporal resolution or (b) imperfect spatial and imperfect temporal resolution, we show an example of a flag received by the user. Given this flag, we show possible locations of leakage, shown on the left for each figure, with probabilities dependent on $\eta$. The blue denotes leakage events that are detected on time, while the red denotes leakage events detected one-time step later. The same potential leakage locations are labeled on the surface code plaquettes, shown on the right in each figure, where the subscripts on $\mathrm{L}$ denote if the leakage is on a data or ancilla qubit.}
    \label{fig:noise_models3}
\end{figure}

Next, we consider the following properties for the erasure checks,
\\
\\
{\bf Check with perfect spatial resolution}, in which a flag received by the user after a two-qubit gate indicates exactly which qubit involved in the gate was leaked. This model is appropriate when each qubit has its own independent erasure check. In this case, the user resets the flagged qubit in a computational basis state, after which the error correction cycle continues from where it was left off. This type of noise on the flagged qubit is accounted for in simulations as a Pauli error drawn from $\mathcal{E}_{\mathrm{L}}=\{I,X,Y,Z\}$ with equal probability~\cite{bennett1997capacities, wu2022erasure}. 
\\
\\
{\bf Check with imperfect spatial resolution}, in which the flag only indicates that there was leakage during the gate but does not indicate which of the two qubits leaked. Imperfect spatial resolution is a property of gates with built-in checks. In this case, the user resets both the qubits participating in the gate and the resulting noise is accounted for in simulations as a Pauli error drawn from $\{I,X,Y,Z\}^{\otimes 2}$ with equal probability~\cite{bennett1997capacities, wu2022erasure}. This noise model is motivated by the noise properties of entangling gates with built-in erasure checks proposed for superconducting dual-rail qubits~\cite{teoh2023dual,tsunoda2023errordetectable} (see Appendix~\ref{sec:builtin}). 

{\bf Check with imperfect temporal resolution}, in which a flag received by the user after a gate could have resulted from a leakage in that gate with probability $\eta$ or from a leakage in an immediately preceding gate on the qubits with probability $1-\eta$. 
We illustrate an example in Figure~\ref{fig:noise_models3}~(a), in which we assume checks have perfect spatial resolution and imperfect temporal resolution. Here, receiving a flag on the data qubit $q$ after the second gate in the circuit means that leakage could have happened on the data qubit during the second gate (in blue) with probability $\eta$, or on the data qubit at the first gate (in red) with probability $1-\eta$. To clarify, the right-hand side of Figure~\ref{fig:noise_models3}~(a) shows the same scenario as the circuit but on the code to make it visually evident that we know the spatial location but not the time of leakage.

In Figure~\ref{fig:noise_models3}~(b), we consider a more complicated scenario with imperfect spatial resolution and imperfect temporal resolution. These checks cannot resolve which qubit has leaked at each gate, so the user assumes the flag could have resulted from the ancilla or the data qubit with equal probability. As a result, both the ancilla qubit $b$ and data qubit $2$ have probability $\eta/2$ to have leaked during the flagged gate, and probability $(1-\eta)/2$ to have leaked during the previous gate. The right-hand side of Figure~\ref{fig:noise_models3}~(b) depicts this same scenario on the code to emphasize that now the spatial location, in addition to the time of the leakage, is unknown.

Checks with imperfect temporal resolution are motivated by the noise properties of superconducting dual-rail qubits. For example, the QND built-in checks proposed in~\cite{teoh2023dual,tsunoda2023errordetectable} fail to detect a certain fraction, say $1-\eta$, of leakage, but this missed leakage is caught by the check of the next gate. This, however, means that if a gate is flagged then with probability $1-\eta$, the leakage actually happened in the qubits during gates that acted on them immediately before the flagged gate. For the built-in checks, $\eta$ can be as high as $0.986$ for reasonable system parameters (see Appendix~\ref{sec:builtin}). As another example, consider false negatives in erasure checks occurring with probability $1-\eta$. Since the erasure checks are QND and the false negative rate is small, both physically realistic assumptions, a missed leakage at one check is almost certainly caught by the next one. This implies that with probability $1-\eta$, the flag raised at one check came due to leakage missed by the preceding check. Our model neglects the possibility of two consecutive false negatives, which is negligible for physically reasonable false negative rates. In a recent transmon-based dual-rail experimental demonstration, an erasure probability of $2.9\%$ was accompanied by a $1.5\%$ false negative probability, giving $\eta=98.6\%$ and negligible double false negatives $2.2\times10^{-4}$~\cite{levine2023demonstrating}. Similarly, a cavity-based implementation quotes an erasure probability of $2.92\%$ with a false negative probability of $3.7\%$~\cite{degraaf2024midcircuit}. These two implementations achieve false negative rates of the same order of magnitude. We calculate experimentally motivated $\eta$ values for the built-in checks used in Table~\ref{table} in Appendix~\ref{sec:builtin}. 

Our numerical results on thresholds and effective error distance for practical noise models and values of $\eta$, $R_\mathrm{e}$ are summarized in Table~\ref{table}. 
Although, in general, temporally imperfect erasure checks $(\eta<1)$ may reduce the effective distance of the code, we find that for the tailored Pauli model, 
 which is realistic for superconducting dual-rail qubits, the distance is not reduced. Additionally, for realistic parameters in Table~\ref{table}, the threshold for built-in erasure checks is $4.23\pm0.02\%$, which is well above the Pauli threshold $\sim1\%$ and only slightly worse than the perfect erasure detection case, $5.56\pm0.03\%$.
Therefore, with a tailored Pauli model, we find that it is possible to achieve a higher threshold and higher effective distance for a $d\times d$ surface code compared to qubits with conventional Pauli noise. Importantly, both built-in spatially imperfect checks and explicit spatially perfect checks achieve these improvements. Recall that built-in checks with cavity-based DRs do not lead to an increase in the length of the error correction cycle and require fewer transmons. Specifically, the number of transmons required to achieve the effective distance of $d_\mathrm{eff}$ with built-in checks is $29\%$ of the number of transmons required to reach the same effective distance if they are used as conventional qubits with Pauli noise (Appendix~\ref{app: Accounting no of transmons}). This is because the built-in checks only require one transmon per ancilla dual-rail qubit (Appendix~\ref{sec:builtin}). Thus, we find that the cavity-based DRs with built-in checks are highly desirable for practical hardware-efficient QEC. 

In the following section, we describe in more detail how the noise model affects the performance of the code and how we implement the weighted union-find (UF) decoder. 

\section{Noise and decoding for imperfect erasures }
\label{sec: Noise and decoding when erasure detection is imperfect}

Irrespective of the spatial resolution, the main impact of imperfect temporal resolution is that a single leakage error can spread to three qubits. To see this, consider the examples illustrated in Figure~\ref{fig:unrot_codedist_reduction}~(a,b). In these examples, a qubit labeled $A$ leaks during a two-qubit gate with another qubit labeled $B$ at time step $T_k$ with probability $p_e$. Then a Pauli error is introduced on qubit $B$ according to the general $\mathcal{E}_{\mathrm{gen.}}$ or tailored $\mathcal{E}_{\mathrm{tail.}}$ noise models. With probability $\eta$ this leakage gets detected by the erasure check corresponding to the gate between $A$ and $B$ at $T_k$, but with probability $1-\eta$ it will get detected by the erasure check corresponding to the gate applied to $A$ at time $T_{k+1}$. Let us label the qubit that $A$ interacts with at $T_{k+1}$ as $C$. Since $C$ has now interacted with a leaked qubit, it also suffers from a Pauli error. Thus, a single error on $A$ has led to an error on two more qubits $B$ and $C$. This can lead to a reduction in the effective distance.
This effective distance reduction occurs when $B$ and $C$ are along a minimum-weight logical operator and both receive an error of the same type as the action of the logical operator on qubits $B$ and $C$.

\subsection{Noise and the effective distance}
\label{sec: noise and the effective distance}
\begin{figure*}
    \centering
    \includegraphics[width=\textwidth]{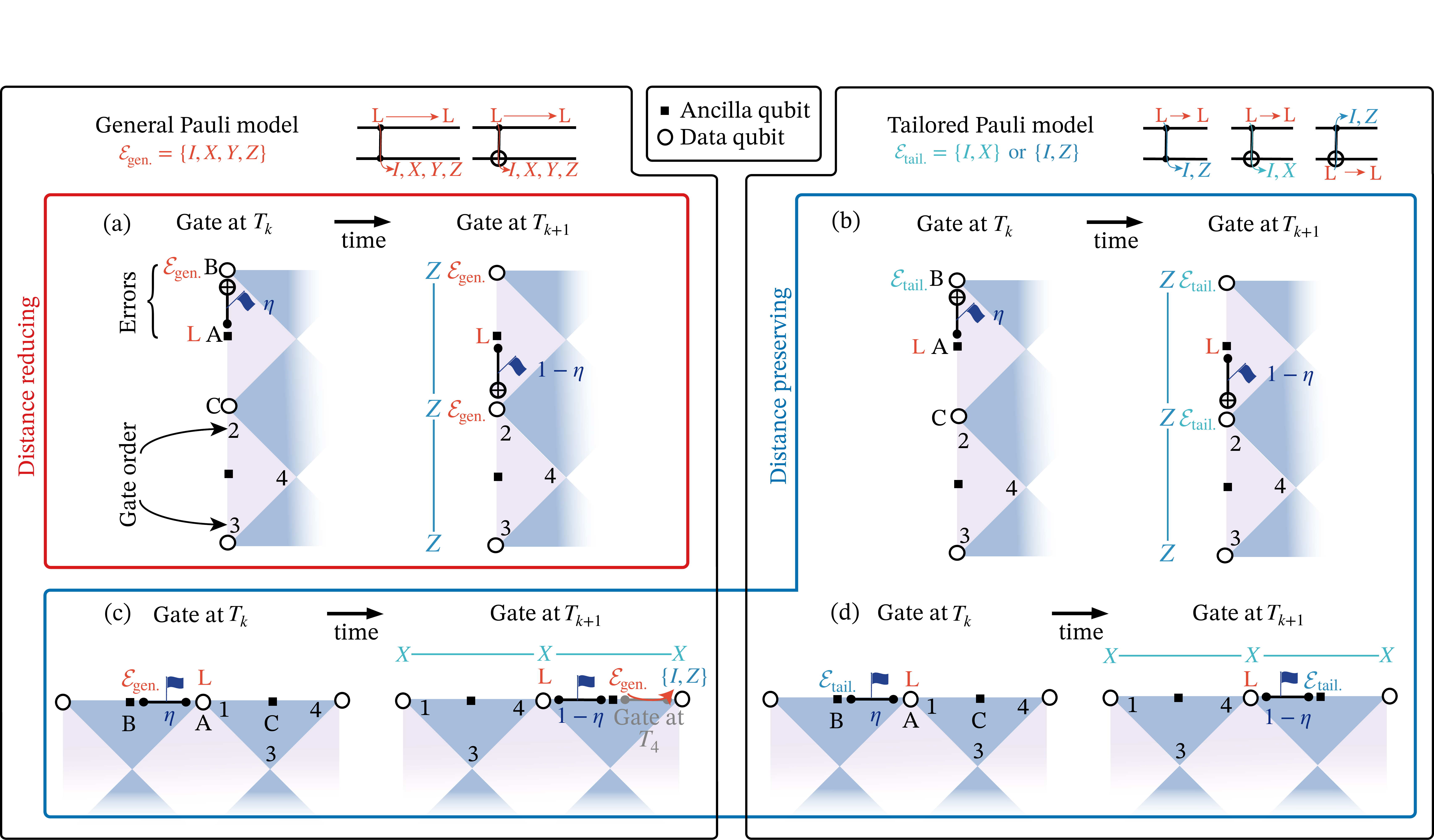}
    \caption{(a) General Pauli model. Data qubits (open circles) and ancilla qubits (squares) along the $Z_\mathrm{L}$. The numbers indicate the order of two-qubit gates, which apply to every stabilizer. Here, we examine a leakage error on the ancilla qubit $A$ during a CX with data qubit $B$ at time $T_k$. With probability $1-\eta$, this leakage is flagged at time $T_{k+1}$. Due to this delayed detection, two data qubits along the $Z_\mathrm{L}$, $B$ and $C$, have errors chosen from $\mathcal{E}_\mathrm{gen.}=\{I,X,Y,Z\}$, thereby reducing the effective distance to a $Z_{\mathrm{L}}$ error. In (b), we consider the same leakage location, but now under the tailored Pauli model. Here, $B$ and $C$ receive faults from $\mathcal{E}_\mathrm{tail.}=\{I,X\}$. These errors do not decrease the effective distance to a $Z_\mathrm{L}$ error. (c) Despite a general Pauli model, the $X_\mathrm{L}$ error distance is preserved due to the scheduling of gates, in contrast to the reduced $Z_\mathrm{L}$ error distance. Although we show one example of data qubit leakage in (c), no other leakage location or time will lead to effective $X_\mathrm{L}$ distance reduction. In (d), we show that the effective distance to a $X_{\mathrm{L}}$ error is also not reduced for the tailored Pauli model. }
    \label{fig:unrot_codedist_reduction}
\end{figure*}

We define the effective error distance $d_{\mathrm{eff}}$ to be the exponent of the physical error rate $p$ in the approximate expression for the logical error rate $p_{\mathrm{L}} \propto p^{d_{\mathrm{eff}}}$ deep in the sub-threshold regime of physical error rates. The effective distance can be roughly understood to be the log probability of the most likely nontrivial logical operator~\cite{dua2024clifford}, which is often, in practice, the minimum number of errors needed to create a logical error. 
As shown in Figure~\ref{fig:unrot_codedist_reduction}~(a), delayed detection can result in two $Z$ errors on $B$ and $C$, which lie in the support of a $Z$-type logical operator. Since one imperfect erasure corrupts two data qubits with a $Z$ error along the boundary, it takes roughly half the number of physical errors to create the most probable $Z_\mathrm{L}$ error. In fact, we expect the $Z_\mathrm{L}$ effective distance to reduce by half and approach $(d+1)/2$ for odd codes, which is the same as the effective distance for pure Pauli noise. 

In contrast to the general Pauli model, the effective distance to a $Z_\mathrm{L}$ error for the tailored Pauli model should remain at $d$. An illustrative example is given in Figure~\ref{fig:unrot_codedist_reduction}~(b), in which data qubits $B$ and $C$ have errors chosen from $\mathcal{E}_{\mathrm{tail.}}=\{I,X\}$, which importantly does not contain $Z$. Therefore, it is impossible for this singular leakage error to lead to multiple $Z$ errors along $Z_\mathrm{L}$. While we show a specific example of ancilla qubit leakage in Figure~\ref{fig:unrot_codedist_reduction} (b), one may repeat this exercise with leakage at other circuit locations and qubits and find that no possible leakage event results in effective distance reduction to a $Z_\mathrm{L}$ error for the tailored Pauli model.

It turns out that effective distance reduction does not depend on whether checks are spatially perfect or imperfect. To elaborate, spatially perfect checks only reinitialize qubit $A$ upon detection, whereas spatially imperfect checks reinitialize both qubits $A$ and $C$. This means if checks are spatially imperfect, qubit $C$ will have errors from $\mathcal{E}_{\mathrm{L}}=\{I,X,Y,Z\}$, but errors on $B$ are unchanged regardless of spatial resolution. This fact, and that the effective distance requires errors on both $B$ and $C$ to have support on the same logical operator means that the effective distance is not dependent on spatial resolution.

Finally, in Figure~\ref{fig:unrot_codedist_reduction}~(c,d), we illustrate that due to the scheduling of gates along the $X_\mathrm{L}$ boundary, one delayed detection does not result in two data qubit errors along $X_\mathrm{L}$. Therefore the effective error distance to an $X_\mathrm{L}$ error is not reduced no matter the noise model. 
For example, let us consider the general Pauli model in Figure~\ref{fig:unrot_codedist_reduction} (c). Leakage on a data qubit $A$ depolarizes two ancilla qubits, $B$ and $C$. As per the usual syndrome measurement procedure, ancilla qubit $B$ is reset before the error can propagate, so let us focus on qubit $C$. Depolarizing noise on qubit $C$ may propagate to a $Z$ error on the right data qubit, however, this $Z$ error cannot reduce the effective distance to an $X_\mathrm{L}$ error. One may go through a similar exercise for leakage on the ancilla qubits, and find that $X$ errors will never appear on two qubits along the $X_\mathrm{L}$ boundary.

We also study the effective distance of the rotated surface code in Appendix~\ref{app:rotated codes}. Unlike the unrotated code, the rotated code's effective distance is reduced along both $X_\mathrm{L}$ and $Z_\mathrm{L}$ under the general Pauli model. However, under the tailored Pauli model, the effective error distance to both an $X_\mathrm{L}$ and $Z_\mathrm{L}$ error is preserved because the pattern of gates in the measurement cycle of the rotated code ensures that the leakage-induced correlated $X$ ($Z$) errors occur on qubits which do not lie along $X_\mathrm{L}$ ($Z_\mathrm{L}$). 

Next, we consider the correlated errors that arise due to imperfect temporal resolution of checks with both perfect and imperfect spatial resolution. We will also describe how the decoder is calibrated for these correlated errors.

\subsection{Noise and decoding: perfect spatial, imperfect temporal resolution}

For all numerical results in this work, we use Monte Carlo simulations of stochastic circuit-level noise and a weighted UF decoder, which has been shown to be effective when both Pauli and erasure noise are present~\cite{delfosse2020linear,delfosse2021almost,huang2020faulttolerant,wu2022erasure,sahay2023high}. The decoder relies on constructing a graph with a vertex at each spacetime syndrome location. A weighted edge is placed between vertices that can be excited by a single circuit fault, and the weight reflects the probability of such a fault. The higher the probability of the fault, the lower the edge weight.

As we saw previously, due to imperfect temporal resolution, a single fault can cause errors on up to three qubits whose locations are exactly known when spatial resolution is perfect. This information is used to calibrate the decoder. Specifically, if a qubit is flagged at time $T_{k+1}$, then this qubit is reset. Irrespective of when the leakage happened, the qubit that interacted with the leaked one at $T_{k+1}$ suffers leakage-induced Pauli noise. Thus, the faults in the gate at $T_{k+1}$ correspond to errors on the leaked qubit drawn from $\mathcal{E}_{\mathrm{L}}$ and those on the qubit that it interacted with are drawn from $\mathcal{E}_\mathrm{gen.}$ {or $\mathcal{E}_\mathrm{tail.}$}. These circuit faults are used to define the corresponding edges of the decoding graph in the standard way. Importantly, the decoder needs to account for the fact that with probability $1-\eta$ the leakage in $A$ could have occurred in the gate that was applied to it at $T_k$. Thus, there could have been additional possible faults corresponding to errors on the leaked qubit drawn from $\mathcal{E}_{\mathrm{L}}$ with probability $(1-\eta)/4$ and those on the qubit that $A$ interacted with at $T_k$, drawn from $\mathcal{E}_\mathrm{gen.}$ or $\mathcal{E}_\mathrm{tail.}$ with probability $(1-\eta)/4$ or $(1-\eta)/2$, respectively. These probabilities are then used to define the edge weights of the decoding graph in the standard way. If no leakage is detected at $T_{k+1}$ then the possible Pauli faults drawn from $\mathcal{E}_p$ with equal probability $p_\mathrm{p}/15$ are accounted for by defining the edges of the decoding graph in the standard way. For details on these weights, refer to Figure~\ref{fig:strong_weak_eras} in Appendix~\ref{sec: decoding details}.

\subsection{Imperfect spatial, imperfect temporal resolution}

In case of imperfect spatial resolution, the flag only indicates an erasure during the gate and thus both qubits, say $A$ and $C$ in Figure~\ref{fig:unrot_codedist_reduction}, participating in the gate are reset if flagged. If a gate is flagged at $T_{k+1}$, then both qubits participating in the gate are reset, and the gate faults correspond to errors on each qubit from $\mathcal{E}_{\mathrm{L}}$ with equal probability $1/16$. These circuit faults are used to define the corresponding edges of the decoding graph in the standard way. Additionally, the decoder must account for the fact that with probability $1-\eta$, $A$ or $C$ could have leaked during a gate that was previously applied at $T_{k}$. Thus, there could have been additional possible faults corresponding to errors on $A$ and $C$, drawn from $\mathcal{E}_{\mathrm{L}}$ with probability $(1-\eta)/4$ and those on the qubits that they interacted with at $T_k$ drawn from $\mathcal{E}_\mathrm{gen.}$ or $\mathcal{E}_\mathrm{tail.}$ with probability $(1-\eta)/4$ or $(1-\eta)/2$ respectively. These probability distributions are then used to define the edge weights of the decoding graph in the standard way (Appendix~\ref{sec: decoding details}). Finally, if no leakage is detected at $T_{k+1}$ then the possible Pauli faults drawn from $\mathcal{E}_p$ with equal probability $p_\mathrm{p}/15$ are accounted for by defining the edges of the decoding graph in the standard way. For details on these weights, we refer to Figure~\ref{fig:strong_weak_eras} in the Appendix~\ref{sec: decoding details}.

\section{Results}

\subsection{Impact on effective error distance}
\label{sec: effective error distance}
\begin{figure}
    \centering
    \includegraphics[width=\columnwidth]{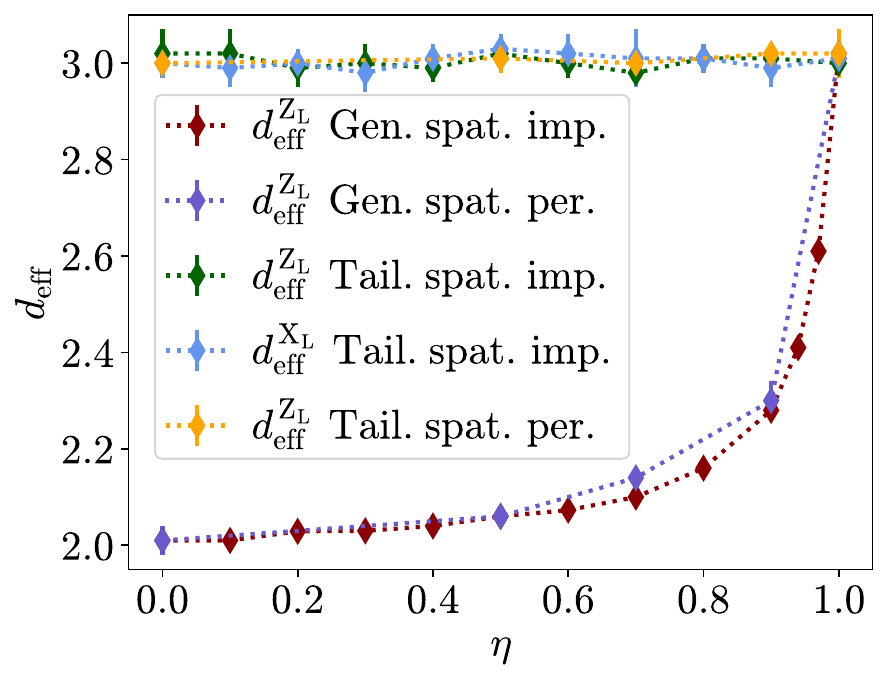}
    \caption{Effective error distance $d_{\mathrm{eff}}$ at $\re=1$ with varying temporal resolution of checks ($\eta$). Red (and purple) diamonds show $d_\mathrm{eff}$ to a $Z_{\mathrm{L}}$ error for the general Pauli model with spatially imperfect (and perfect) checks. The dotted lines connecting the diamonds are a guide to the eye. $d_{\mathrm{eff}}$ to a $Z_{\mathrm{L}}$ error is shown for the tailored Pauli model with spatially perfect checks (orange) and imperfect checks (green). Light blue shows $d_{\mathrm{eff}}$ to an $X_{\mathrm{L}}$ error for the tailored Pauli model with imperfect checks.}
    \label{fig:codedist_v_eta}
\end{figure}

We numerically confirm that the general and the tailored Pauli models have different consequences on the effective error distance, $d_{\mathrm{eff}}$. In Figure~\ref{fig:codedist_v_eta}, we plot the effective error distance to a logical error by fixing the code size ($3\times3$) and varying the temporal resolution $\eta$, the Pauli model, and the spatial resolution. We assume physical errors are exclusively erasures ($R_\mathrm{e}=1$), thus, the effective error distance should be $3$ when erasure detection is perfect. We extract $d_{\mathrm{eff}}$ by plotting the logical error rate for varying physical error rates that are significantly below the threshold ($p \sim 0.1p_{\mathrm{th}}$). At this regime of low $p$, we approximate $p_{\mathrm{L}}\propto p^{d_{\mathrm{eff}}}$ and plot $\log(p_{\mathrm{L}})$ v. $\log(p)$. The slope of this line is the effective error distance. 

The light blue diamonds in Figure~\ref{fig:codedist_v_eta} are numerically extracted effective distances to an $X_{\mathrm{L}}$ error under the general Pauli model with spatially imperfect checks. The dots connecting the data points in diamonds are a guide to the eye. As expected, we notice the effective distance to an $X_{\mathrm{L}}$ error is not reduced and remains at $3$ regardless of the temporal resolution of checks. Since the remaining noise models are more restrictive than the general Pauli model with spatially imperfect checks (e.g. tailored, spatially resolved checks), we do not expect the effective distance to an $X_{\mathrm{L}}$ error in these models to be worse than $3$. Furthermore, by virtue of examining a distance $3$ code, the effective distance will never be greater than $3$.

The remaining points in the figure illustrate the effective distance to a $Z_{\mathrm{L}}$ error. Now, let us examine how the effective error distance changes with different leakage-induced Pauli models. The red curve in Figure~\ref{fig:codedist_v_eta} examines the effective error distance to a $Z_{\mathrm{L}}$ error under the general Pauli model with spatially imperfect checks. As predicted in Section~\ref{sec: noise and the effective distance}, the effective distance quickly approaches $2$ as the temporal detection accuracy gets worse ($\eta\rightarrow0$). This is the Pauli noise scaling for a $d=3$ code ($p_{\mathrm{L}}\propto p^{(d+1)/2}=p^{2}$). Therefore, with delayed erasure detection and the general Pauli model, the $Z_\mathrm{L}$ error rate no longer scales advantageously compared to Pauli errors. We confirm that the Pauli scaling $p^{(d+1)/2}$ is reproduced for larger lattice sizes in Figure~\ref{fig:codedist_vary_size} of Appendix.~\ref{app: lattice sizes}. 

Conversely, we observe that in the tailored Pauli case, with spatially imperfect checks (green) and spatially perfect checks (orange), the effective error distance remains at $3$ (within error bars) regardless of the spatial resolution of checks or how temporally imperfect the detection is. This means the logical error rate of both $X_\mathrm{L}$ and $Z_\mathrm{L}$ remains the scaling for perfect erasures, $p^d$, for the tailored Pauli model, as predicted in Section~\ref{sec: noise and the effective distance}. 

In general, we see that the effective error distance is unaffected by the spatial resolution of checks. As we just observed, the tailored Pauli model has distance $d$ regardless of the spatial resolution of checks. Furthermore, when we examine the effective distance to a $Z_{\mathrm{L}}$ error for the general Pauli model with spatially perfect checks (purple), there is no difference in $d_{\mathrm{eff}}$ to spatially imperfect checks (red) within error bars. This means that in terms of effective error distance, there is no advantage to engineering erasure checks which are spatially resolving. Finally, we numerically show that for the experimentally relevant tailored Pauli model, the effective distance to an $X_\mathrm{L}$ (light blue) or $Z_\mathrm{L}$ (green) error does not drop due to delayed erasure detection.

\subsection{Impact on threshold}
\label{sec: Impact on Threshold}
Next, we numerically examine the consequences of imperfect detection resolution on the threshold. All thresholds are obtained with lattice sizes $d=9,11,13,15$, and fitted by using the quadratic expansion method used in~\cite{WangChenyang2003}.

\begin{figure}[!htb]
    \centering
    \includegraphics[width= \columnwidth]{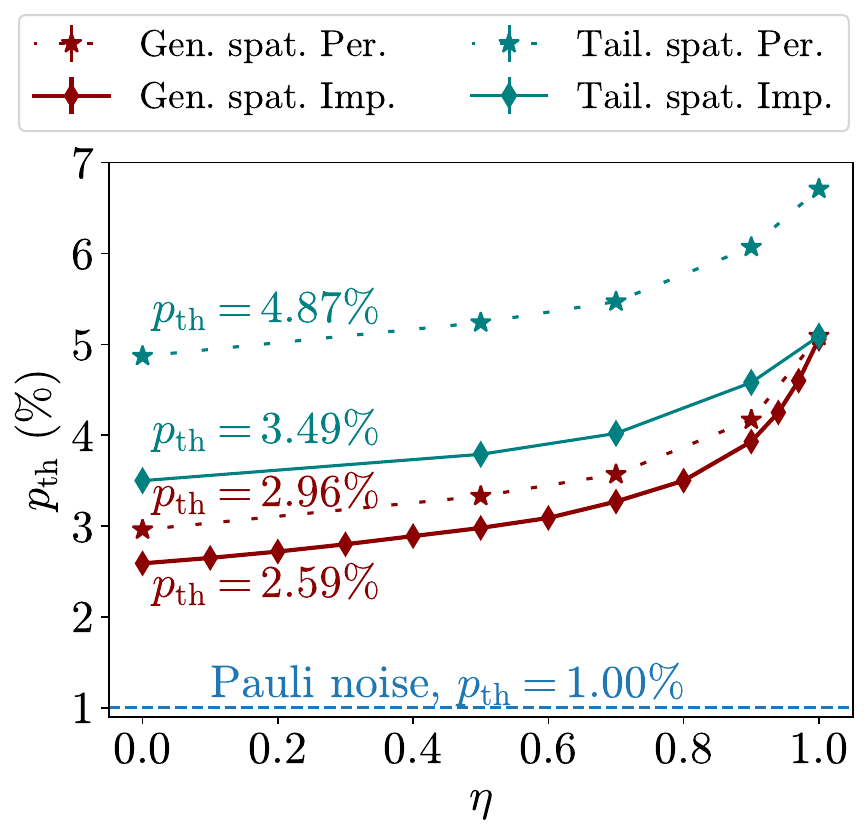}
    \caption{Threshold for varying temporal resolution ($\eta$) at $R_e=1$ for four noise models: the general Pauli model with perfect (red stars) and imperfect (red diamonds) spatial resolution, and the tailored Pauli model with perfect (teal stars) and imperfect (teal diamonds) spatial resolution. The connecting lines between points are a guide to the eye. The threshold for Pauli noise after every two-qubit gate is marked at the dotted blue line.}
    \label{fig:re=1_thres_v_eta}
\end{figure}

We continue to consider the case of only erasure noise ($R_\mathrm{e}=1$) in Figure~\ref{fig:re=1_thres_v_eta}. For comparison purposes, we plot the threshold for pure Pauli noise, $1.00 \pm .01\%$, marked in the dotted blue line. We plot the threshold against $\eta$ for the general and tailored Pauli models with and without spatially perfect checks. For all four noise models, the threshold drops the quickest in the regime of near-perfect temporal detection ($\eta \sim 1$). Thus, if the detection mechanism is only marginally temporally inaccurate, it pays off in threshold to improve the temporal detection accuracy slightly.

We observe that the thresholds for the tailored \& spatially imperfect, general \& spatially perfect, and general \& spatially imperfect noise models all approach the threshold for perfect erasures when $\eta = 1$. This is an expected consequence since at exactly $\eta=1$ all three noise models are equivalent. To elaborate, for spatially imperfect checks (general and tailored), both qubits participating in the gate are reset when the gate is flagged, and if detection is on time, these are the only faults on the qubits. Thus, when $\eta\rightarrow 1$, most of the modified edges on the decoding graph correspond to faults chosen from $\mathcal{E}_{\mathrm{L}}$ at $T_{k+1}$ on each qubit given a flagged gate at $T_{k+1}$. When all detection events are on time, the edges that are modified in the general spatially perfect model are the same as the edges modified for the general spatially imperfect checks. This is because the un-leaked qubit participating in the flagged gate has faults drawn from $\mathcal{E}_{\mathrm{gen.}}$, which contain the same errors as $\mathcal{E}_{\mathrm{L}}$. Since the edges that are erased when $\eta=1$ are the same for these three models, we expect that the thresholds should converge to the same value when all detection events are on time.

On the other hand, only the tailored Pauli model with spatially perfect checks does not converge to the same threshold value at $\eta=1$ as the rest of the noise models. Given a flagged gate with on-time detection, the flagged qubit has faults chosen from $\mathcal{E}_{\mathrm{L}}$ and the unleaked qubit has an error from the set $\mathcal{E}_{\mathrm{tail.}}$. Due to the restricted error set of $\mathcal{E}_{\mathrm{tail.}}$, we can know if the unleaked qubit has a $Z$ or $X$ error depending on the type of gate applied (CX or CZ). This results in an erasure with extra information, where we know not only the location of the error but also what type ($X$ or $Z$) it is. On the decoding graph, this results in fewer edges that are modified/set to a lower weight. In the language of percolation theory and its relationship to the erasure error threshold, when fewer edges are removed with probability $p$, the threshold error rate is higher~\cite{stace2009thresholds, Barrettstace2010}.

The convergence in thresholds for temporally-accurate checks results in interesting consequences. For example, in the general Pauli model, the difference in threshold between spatially imperfect (red diamonds) and perfect (red stars) checks is small in the regime of $\eta\rightarrow 1$. This means that the cost of engineering checks that can resolve which qubit is leaked at every two-qubit gate is hardly going to increase the threshold. Whereas in the tailored Pauli model case, the difference in threshold between perfect (teal stars) and imperfect (teal diamonds) spatial resolution in this regime is over $1\%$.

Finally, as late detection is more frequent $(\eta\rightarrow0)$, the threshold approaches $2.59\pm 0.01\%$, $2.96 \pm 0.02\%$, $3.49 \pm 0.02\%$ and $4.87 \pm 0.05\%$ for the general Pauli model with spatially imperfect and perfect checks, and the tailored Pauli model with spatially imperfect and perfect checks respectively. In all four noise models, this is well above the Pauli noise threshold of $1.00 \pm 0.01\%$. This means even in the worst-case scenario where most detection cannot be relied on to be temporally accurate, erasures with delayed detection will always have a higher threshold than for Pauli noise. 
This performance at $\eta=0$ is interesting for erasure checks when real-time classical processing of the measurement outcome takes about as long as a two-qubit gate. To eliminate this extraneous time between gates, one may decide to proceed with the next two-qubit gate before knowing the outcome of the erasure check. This means that every result reflects the leakage of qubits one-time step prior.

\begin{figure}
    \centering
    \includegraphics[width=0.9\columnwidth]{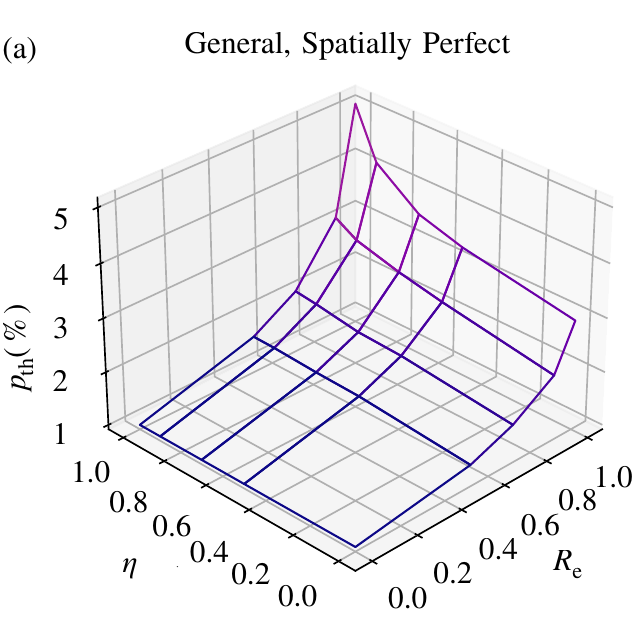}
    \includegraphics[width=0.9\columnwidth]{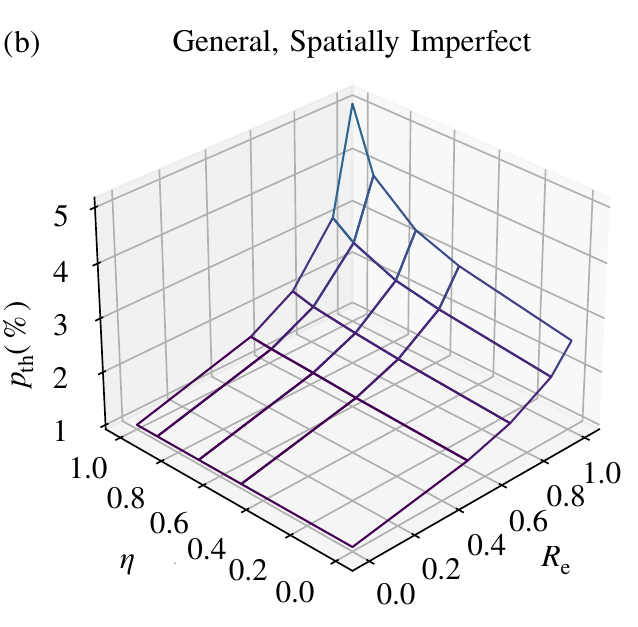}
    \caption{(a) For spatially perfect checks and the general Pauli model, we plot threshold (Z axis) parameterized over $\eta$ and $\re$. (b) Similarly, we repeat this for spatially imperfect checks with the same leakage-induced Pauli model. The threshold values in these plots are listed in table-form in Appendix~\ref{app:3d threshold plots}.}
    \label{fig:phase_plots_general}
\end{figure}

\begin{figure}
    \centering
    \includegraphics[width=0.9\columnwidth]{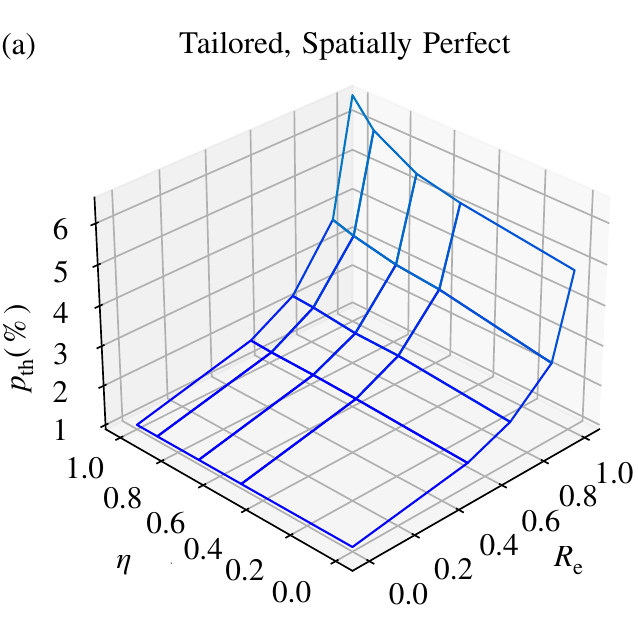}
    \includegraphics[width=0.9\columnwidth]{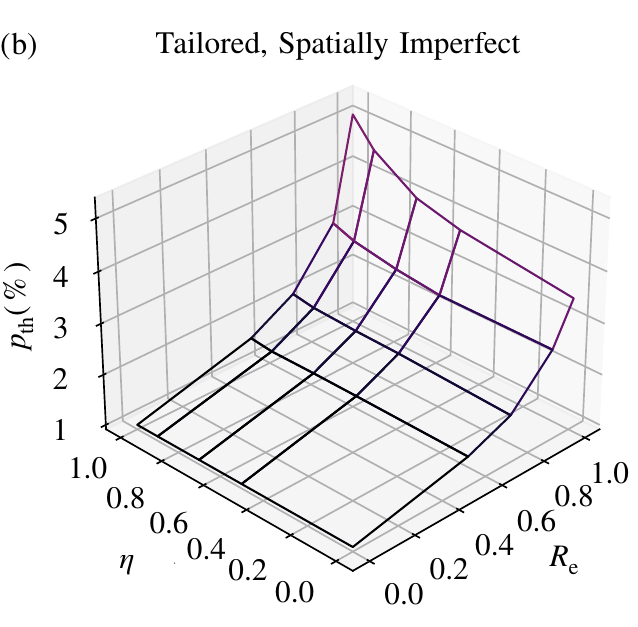}
    \caption{(a) For spatially perfect checks and the tailored Pauli model, we plot threshold (Z axis) parameterized over $\eta$ and $\re$. (b) Similarly, we repeat this for spatially imperfect checks with the same leakage-induced Pauli model. The threshold values in these plots are listed in table-form in Appendix~\ref{app:3d threshold plots}.}
    \label{fig:phase_plots_tailored}
\end{figure}

We have so far discussed what happens when the errors are only erasures and we neglect any amount of Pauli errors. Now, we will show a 3D plot of the threshold parameterized over two variables, the temporal resolution $\eta$ and erasure fraction $R_\mathrm{e}$ for the general Pauli model in Figure~\ref{fig:phase_plots_general}~(a,b) and the tailored Pauli model Figure~\ref{fig:phase_plots_tailored}~(a,b).

The thresholds all converge to $1.00 \pm 0.01\% $ when there is only Pauli noise ($R_e = 0$). We see the largest variation in thresholds among the four noise models when the erasure fraction is high because the thresholds are dominated by erasures and are susceptible to variations in the accuracy of erasure detection. 

\section{Discussion}
In this work, we have analyzed the surface code parameters of threshold and effective error distance for erasure qubits with imperfect erasure checks. We find that erasure qubits outperform qubits with dominant Pauli noise on these metrics 
even when the erasure checks are imperfect, particularly if the leakage-induced noise has an inherent structure like that in the tailored Pauli model. Thus, our work firstly highlights the continued advantages of erasure qubits, and further emphasizes the importance of measuring the detailed noise model in experiments which can be non-trivial and may require the development of new techniques~\cite{claes2023estimating}. 

In this work, we have assumed that all leakage errors are detected within two time steps. While this is a very good approximation, a reduced amount of leakage can persist for longer times. In fact, if leakage occurred at a certain time and the leakage detection efficiency is $\eta$, then $\eta^k$ fraction of leakage remains after $k$ time steps. The residual leakage may limit the performance of the code unless checked by a leakage-reduction unit or by inherent control techniques. We leave this analysis for future work.

Finally, our work motivates further analysis of the trade-offs between simplifying the erasure checks and code parameters. One simple approach to reduce the time-cost of erasure checks is to perform them more infrequently, such as, once per stabilizer measurement. However, this would result in a single leakage event persisting for the whole stabilizer measurement cycle which could degrade the threshold and effective distance~\cite{baileyinprep}. It would be fruitful to identify practical gates and leakage-induced Pauli models which do not reduce the effective distance even with infrequent erasure checks.

\section{Acknowledgements}
This research was supported
by the U.S. Army Research Office (ARO) under grant W911NF-23-1-0051. KC would like to thank Yue Wu, Aleksander Kubica, and Bailey Guo for the useful discussions. KC and SS also thank Akshay Kootandavida for discussions on experimental implementations of the dual-rail.

\textit{Note-- A related, independent work~\cite{baileyinprep} examines alternative erasure check schedules to relax check requirements on the performance of the surface code.}

\bibliography{imp_eras}

\clearpage
\appendix

\section{Gate with built-in erasure check}
\label{sec:builtin}
In this gate, which is described in more details in the following sections, the gate and erasure check are simultaneously implemented by coupling the dual-rail with a transmon. At the end of this gate, the transmon is measured. Depending on the measured state, one can deduce if there was an error during the gate, flagging the gate~\cite{teoh2023dual,tsunoda2023errordetectable}. The transmon raises two kinds of flags: one due to a relaxation event ($T_1$ event) in the transmon itself and the other due to a dephasing event in the transmon or due to a relaxation/photon loss event in the dual-rail. There are a few important points to note. Firstly, the transmon flag cannot differentiate between a dephasing event and an erasure in the dual-rails. Secondly,  it cannot resolve between erasures in the two dual-rail qubits participating in the gate. Thirdly, once a transmon $T_1$ error is flagged, both the dual-rails must be reset as they may have incurred an unknown leakage error. 
Thus, the above observations imply that the gate has imperfect spatial resolution because as soon as any flag is seen, there is an uncertainty in the type of error and both dual-rails must be reset. Finally, the transmon flags erasure or photon loss in the dual-rail only with $50\%$ probability. Thus, it also has an imperfect temporal resolution which depends on the rate of single-photon loss in the dual-rails ($\kappa$), and the rates of dephasing $1/T_\phi$ and relaxation in the transmon $1/T_1$ (note that here $T_\phi$ and $T_1$ are that of the $f$-level of the transmon), $\eta=(\kappa+1/T_1+1/T_{\phi})/((2\kappa+1/T_1+1/T_{\phi}))$. For conservative experimental parameters of $T_1=30$ $\mu$s, $T_\phi=25$ $\mu$s and $\kappa=1$ ms$^{-1}$ we have $\eta\sim 0.986$.

\begin{figure}
    \centering
    \includegraphics[width=\columnwidth]{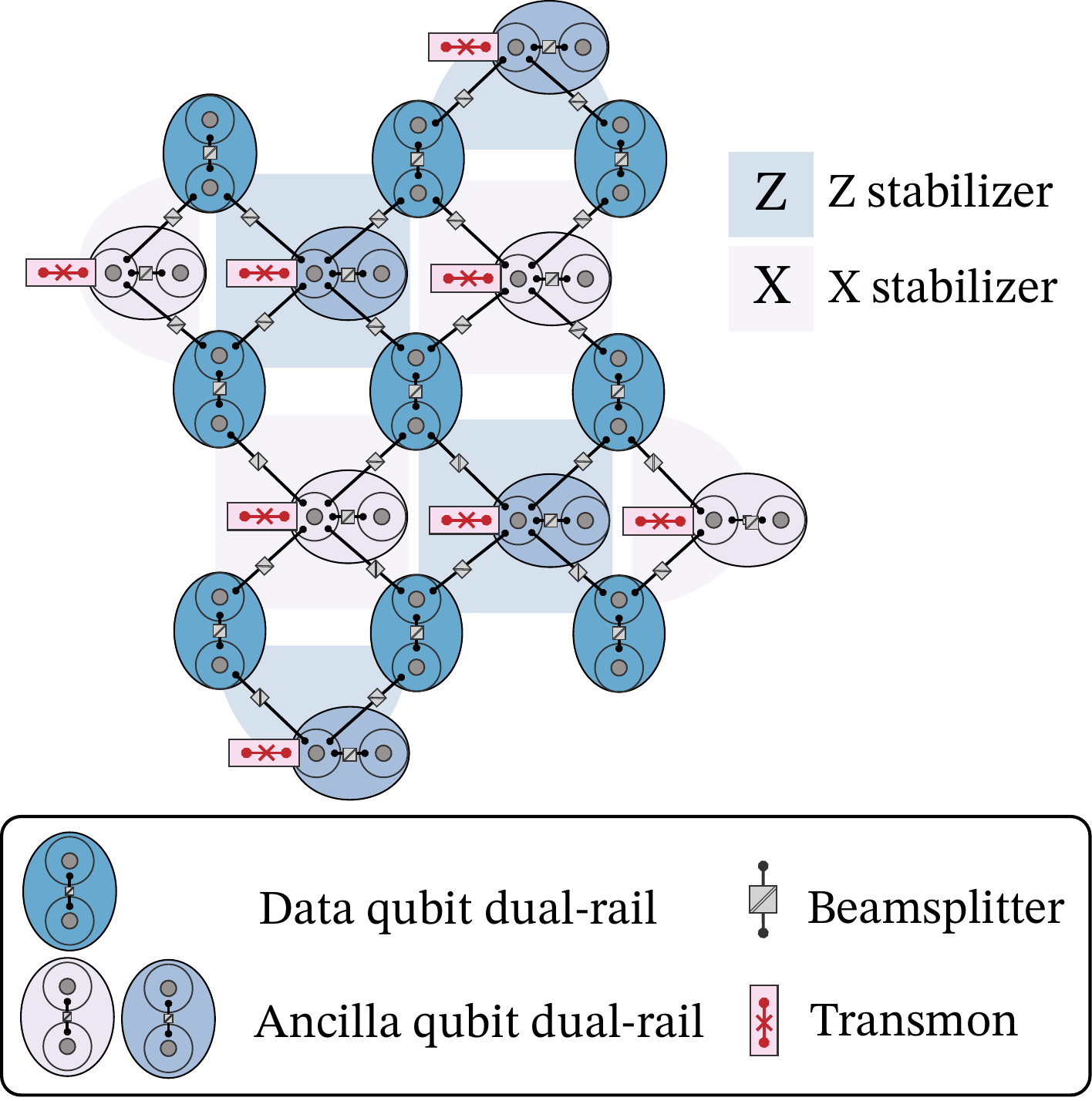}
    \caption{Potential layout of transmons and cavities in a surface code implemented with cavity-based dual-rail qubits with built-in erasure checks. Instead of explicit erasure checks on each qubit, which would require an extra transmon and readout line on every physical qubit, the built-in nature of these checks halves the number of these ancillary transmons.}
    \label{fig:dr surface code}
\end{figure}

Due to the built-in nature of this erasure check, in principle, one only needs one ancilla transmon for every ancilla dual-rail qubit on the surface code. We illustrate a potential layout of hardware elements in Figure~\ref{fig:dr surface code} of a $3\times 3$ rotated code. This would only require 8 readout lines (one per ancilla transmon), in comparison to 17 readout lines for explicit erasure checks on every physical qubit. More details on hardware costs of different architectures are in Appendix~\ref{app: Accounting no of transmons}.

\section{Overview of the gates with and without built in checks}
\begin{figure}
    \centering
    \includegraphics[width=\columnwidth]{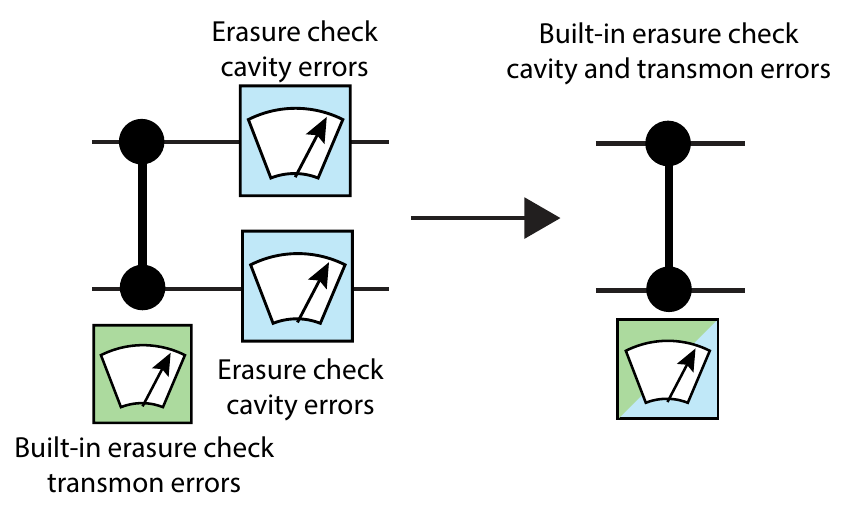}
    \caption{(Left) Typical use case for an entangling gate between two erasure qubits. Erasure checks are performed after the gate on both qubits. In the gate proposal in \cite{teoh2023dual} for dual-rail cavity qubits, the gate comes with `built-in' error detection for transmon ancilla errors which may occur during the gate. However, separate erasure checks are needed after the gate to detect cavity errors. (Right) Our proposed gate modification combines the error detection of transmon and cavity errors into a single measurement (of the transmon). As a consequence of this simplification, we will be unable to detect cavity photon loss errors if they occur mid-way through the gate with 100\% probability, thus motivating our need to study delayed erasure detection.}
    \label{fig:inbuilt summary}
\end{figure}

Now, we describe in more detail two scenarios of erasure checks presented in Figure~\ref{fig:inbuilt summary}. We first consider two-qubit gates as proposed in \cite{teoh2023dual,tsunoda2023errordetectable} followed by explicit erasure checks for cavity leakage. 
The scheme for the two-qubit gate leverages the dispersive interaction between a transmon and one of the cavities to perform a $ZZ(\pi/2)$ maximally entangling gate. Transmon errors, if undetected, propagate onto the cavities as either dephasing or leakage errors. However, detecting a transmon error via transmon readout after the gate allows us to convert these errors to erasures.
One prominent feature of the gate is the ability to read out the state of the transmon at the end of the gate, which allows us to detect whether transmon errors have occurred during the gate. In particular, transmon decay errors are mapped to the $\ket{e}$ state while transmon dephasing errors are mapped to the $\ket{f}$ state. Measuring the $\ket{g}$ state signals that no transmon error has occurred during the gate. If we measure the transmon to be in $\ket{e}$ or $\ket{f}$ after the gate, we must reset both dual-rail cavity qubits. 

Detecting transmon errors allows us to use the dispersive interaction between the transmon and the cavity to perform a dual-rail entangling gate reasonably quickly (on the order $\sim1 \mu s$ gate time) while preventing our Pauli error rate from being limited by transmon decoherence. 

However, it is not sufficient to detect transmon ancilla errors in a dual-rail cavity qubit alone. We must also detect photon loss errors and convert these to erasures (even though they are expected to be 10 times rarer). Otherwise, photon loss will lead to a disastrous accumulation of leakage errors. 

The $ZZ$ gate proposed in \cite{teoh2023dual} actually already has the ability to detect some photon loss errors if they occur during the gate. In particular, a photon loss error dephases the transmon ancilla (by inducing a $Z$ rotation by an angle which depends on the exact time of the photon loss), and so it will be detected in the $\ket{f}$ state with 50\% probability at the end of the gate. 
In order to catch any remaining leakage or leakage, erasure checks must be performed by measuring the joint parity operator of each dual-rail qubit after the gate~\cite{teoh2023dual}.

These erasure checks take time and in turn, expose us to more cavity and transmon errors. Instead, we propose modifying the $ZZ$ gate protocol to have an in-built check for photon loss errors in the cavities, thereby eliminating the need for costly dedicated erasure check operations. This new paradigm is illustrated in Fig \ref{fig:inbuilt summary} and also allows us to place transmon ancillas more sparsely in a dual-rail architecture, in every other physical qubit. 

Our new gate will instead have the following error detection properties:

\begin{enumerate}
    \item Transmon decay during the gate $\rightarrow\ket{e}$
    \item Transmon dephasing during the gate
    $\rightarrow\ket{f}$
    \item Leakage on a dual-rail qubit \textit{during} the gate
    $\rightarrow\ket{f}$ (with $\approx50\%$ probability)
    \item One dual-rail qubit is leaked \textit{before} the gate 
    $\rightarrow \ket{f}$ 
    \item Both dual-rail qubits leaked \textit{before} the gate
    $\rightarrow \ket{g}$
\end{enumerate}
To reiterate, properties $1-3,5$ are inherent to the ZZ gate proposed in \cite{teoh2023dual} while $4.$ results from our proposed modification. Item $5.$ is an undesirable property still present in our modification in that we are unable to reliably detect if both dual-rail qubits are leaked before or during the gate. However, this is a second order error which could be detected with sparsely interspersed erasure check operations. 

\begin{figure}[h!]
    \centering
    \includegraphics[width=\columnwidth]{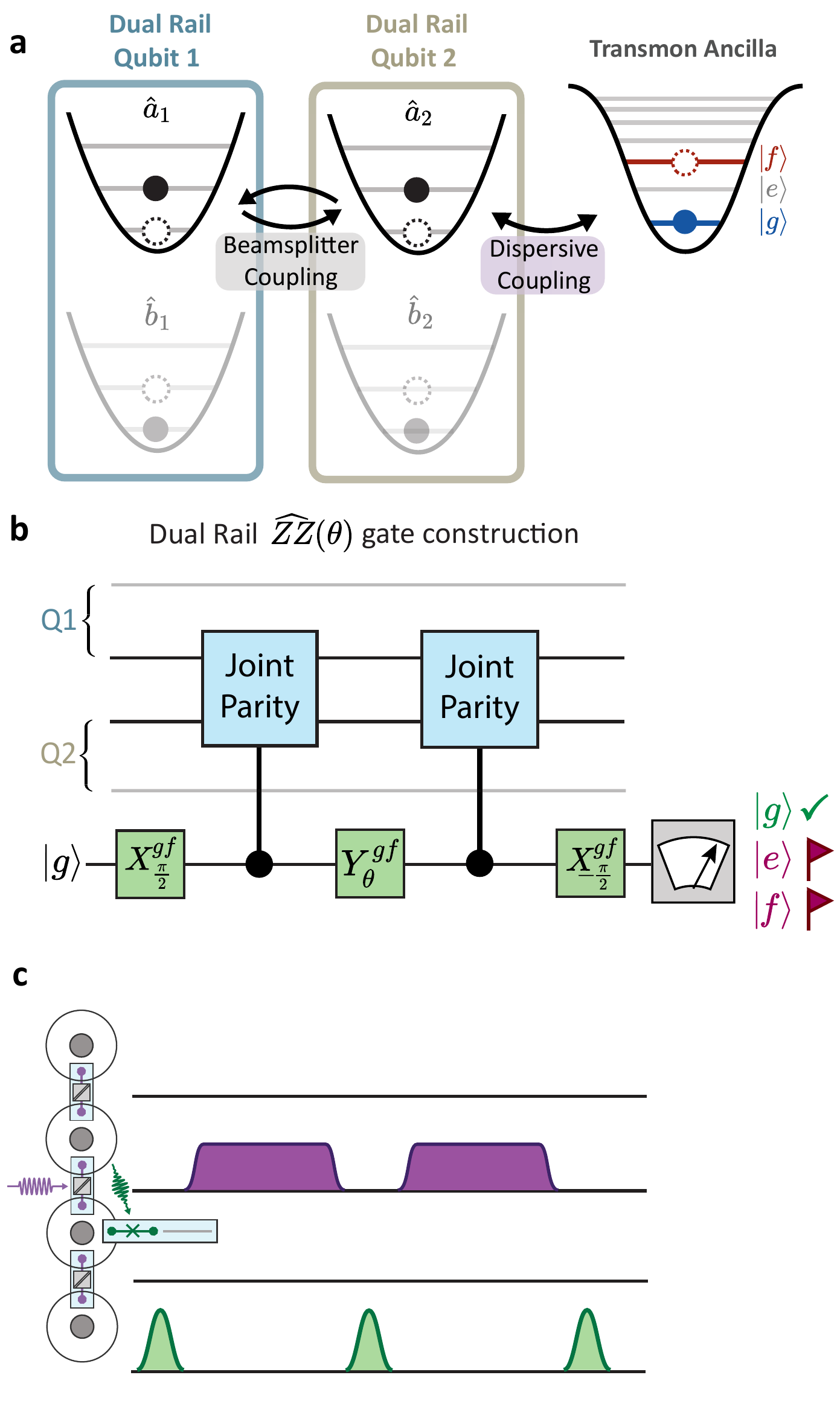}
    \caption{a) Dual-rail ZZ gate as proposed in \cite{teoh2023dual}. With only a parametric beamsplitter between the middle two cavity modes, $(b_1,a_2)$ and a transmon dispersively coupled to only cavity $a_2$, it is possible to perform a $ZZ(\theta)$ entangling gate between the two dual-rail qubits. b) Quantum circuit construction for the dual-rail $ZZ(\theta)$ gate. This construction already has some degree of error-detection built in. Measuring the transmon in $\ket{g}$ at the end of the gate signals no transmon errors have occurred during the gate. By performing transmon ancilla rotations in the $g$-$f$ manifold instead of the $g$-$e$ manifold, we can detect transmon decay errors to first order, which results in measuring the transmon in $\ket{e}$ at the end of the gate. Transmon dephasing errors are detected by measuring the transmon to be in $\ket{f}$. Photon loss errors in either dual-rail qubit during the gate result in a 50-50 probability of measuring $\ket{g}$ or $\ket{f}$ at the end of the gate and hence are only partially error detected. c) Hardware layout and pulse sequence required for the dual-rail $ZZ(\theta)$ gate. The pulse sequence is a qualitative representation of the controls required and typical durations for implementing the gate.}
    \label{fig:ZZ gate original}
\end{figure}
\newpage~\newpage

\section{Details of the gate and leakage detection}
Next, we briefly describe the $ZZ$ gate in \cite{teoh2023dual}. First, we write the four dual-rail basis states in the $Z$ basis as $\{\ket{01,01},\ket{01,10},\ket{10,01},\ket{10,10}$, using the notation $\ket{N_{a_1},N_{b_1},N_{a_2},N_{b_2}}$ 
If we examine the joint-parity of the middle two cavities, $a_2$ and $b_1$ we observe that $\{\ket{01,01},\ket{10,10}\}$ have odd joint-parity while $\{\ket{01,10},\ket{10,01}\}$ have even joint-parity. The $ZZ$ gate puts a relative phase between these two sets of logical states and can be achieved by the circuit in Fig. \ref{fig:ZZ gate original} (a).

It is apparent that if a dual-rail qubit begins in the leakage state $\ket{0,0}$, there is no way to detect this by only measuring the middle two cavities with the transmon ancilla. (e.g. the state $\ket{00}$ in dual-rail $(a_1,b_1)$ looks `identical' to the state $\ket{10}$ if only cavity $b_1$ interacts with the transmon).

\begin{figure}[h!]
    \centering
    \includegraphics[width=\columnwidth]{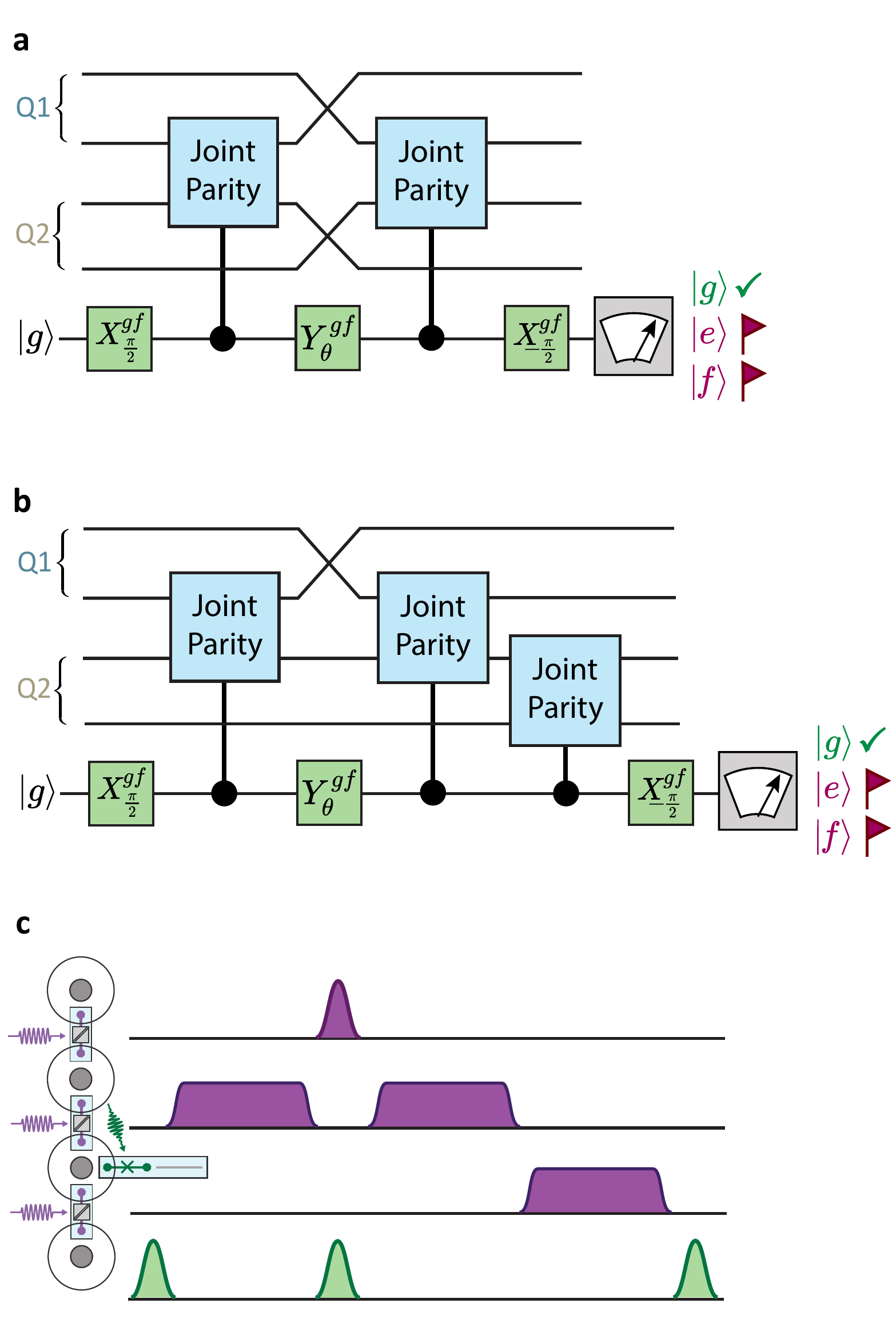}
    \caption{(a) Modified quantum circuit for implementing a dual-rail $ZZ(\theta)$ gate with in-built erasure detection for cavity photon loss errors. Without additional SWAP operations after the gate, the circuit implements the entangling gate $X_1X_2ZZ(\theta)$.  If one out of the two dual-rail qubits begins in a leakage state such as $\ket{00}$ or $\ket{11}$ we will measure the transmon to be in $\ket{f}$ at the end of the gate. This allows us to remove the expensive dedicated erasure check operations from our implementation of the surface code. If photon loss happens mid-way through the gate, we only have a 50\% chance of measuring the transmon in $\ket{f}$, leading to $\eta\approx0.5$ for photon loss erasures when using this gate.  (b) A more realistic proposal which does not rely on performing fast cavity swaps between $(a_2,b_2)$ relative to the dispersive interaction strength. This sequence has the same error-detection properties as the circuit in a) and implements the entangling gate $X_1 ZZ(\theta)$. (c) Qualitative pulse sequence and hardware required to implement the circuit in (b). Compared to the original $ZZ$ gate scheme, our proposal does not require any additional hardware but lengthens the gate duration by around 50\%.}
    \label{fig:ZZ gate improved}
\end{figure}

We can allow the transmon to measure the state of all four cavities during the $ZZ$ gate by simply adding cavity-cavity SWAPs between pairs $(a_1, b_1)$ and $(a_2,b_2)$ in between the control joint-parity unitaries, as shown in Fig. \ref{fig:ZZ gate improved} (a). After the gate, we may optionally perform another set of cavity swaps or merely track this as additional $X$ gates on both dual-rail qubits.

For the transmon to be in $\ket{g}$ at the end of the gate, both controlled joint-parity unitaries must induce the same phase shift (0 or $\pi$) on the ancilla transmon. 

Suppose we began in the state $\ket{00,01}$ before we perform the gate. The first control joint parity unitary will induce 0-phase shift on the transmon ancilla. However, after the swaps, the state in the cavities will be $\ket{00,10}$ for the second control-JP unitary and hence a $\pi$ phase shift ($Z$ operator) on the transmon, and we will measure the transmon to be in $\ket{f}$ instead of $\ket{g}$, thereby detecting this photon loss error. 

In other words, adding these swap operations means we are also measuring if the joint parity of $(b_1,a_2)$ is the same as the joint parity of $(a_1,b_2)$. Crucially, if we are still in the dual-rail codespace, we also do the desired $ZZ$ entangling gate.

In practice, performing the swap on cavity pair $(a_2,b_2)$ while the transmon is still in a superposition is thought to be impractical, because the strength of the dispersive interaction is comparable to the typical strength of the parametric beamsplitter, and so the beamsplitter interaction will be off-resonance the transmon is in the $\ket{f}$ state. (It is possible to perform an 'unconditional' SWAP as shown in \cite{tsunoda2023errordetectable} but this is comparable or longer than a control joint-parity operation).

In comparison, the $(a_1,b_1)$ swap does not have this issue, since any transmon ancillae coupled to these cavities should remain in their ground state throughout the gate. 

The final circuit and pulse sequence we suggest for experimental implementation is shown in Fig. \ref{fig:ZZ gate improved} (b) and c., in which we perform the $(a_1,b_1)$ SWAP halfway through the gate, and add an extra control joint-parity unitary acting on $(a_2,b_2)$. Overall, our modifications will make the gate $\approx 50\%$ longer in duration but now effectively lump the check for transmon errors and erasure checks on the dual-rails into a single transmon measurement, leading to an overall reduction in the time needed to perform a round of stabilizer measurement and few transmon ancilla readout operations.

\section{Accounting for the number of transmons and control lines in different proposed dual-rail architectures}
\label{app: Accounting no of transmons}

Dual-rail qubits are inherently composite systems and hence require more hardware to realize each physical qubit. At the same time, their properties as erasure qubits promise more efficient error correction which has the potential to greatly reduce the number of total hardware elements required to reach some target logical error rate. It is important that we can quantify and estimate the hardware savings we can achieve with dual-rail qubits, which in turn requires quantifying these two affects. 

\begin{figure}[h!]
    \centering
    \includegraphics[width=0.4\columnwidth]{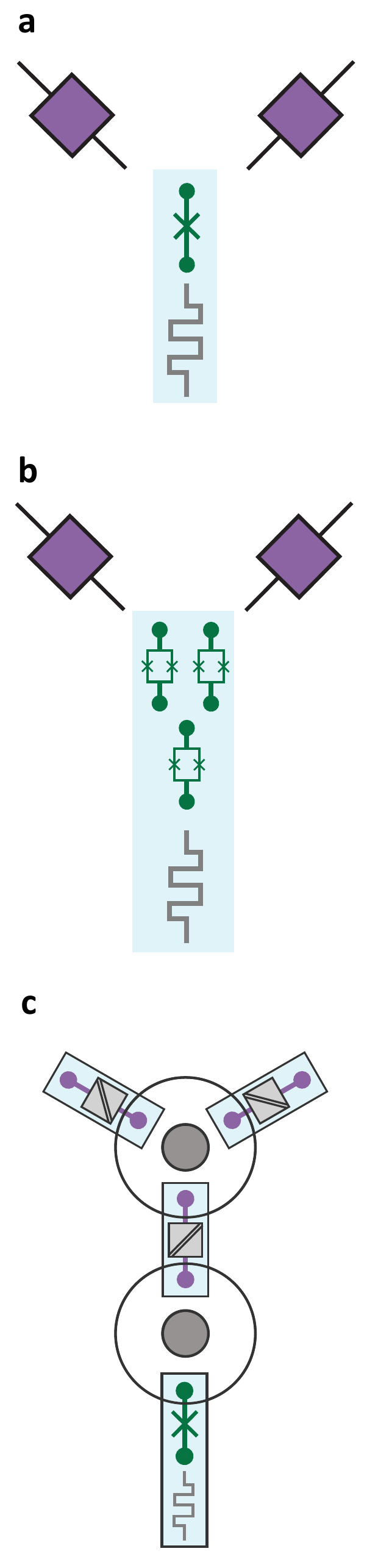}
    \caption{Tile-able unit cells for different superconducting qubit architectures. These unit cells can be tiled in a plane to produce a square lattice of physical qubits with nearest-neighbor connectivity in order to realize a surface code. Counting the number of hardware elements per unit cell is a useful point of reference when comparing different possible qubit architectures. (a) Possible unit cell for transmon qubits (green) with tunable couplers (purple boxes). These tunable couplers are presumed to be transmon-like elements required for entangling gates. Each transmon is also presumed to have its own readout resonator (grey). (b) Possible unit cell for dual-rail transmon qubits as demonstrated in \cite{levine2023demonstrating} (in another version, each transmon may have its own dedicated readout resonator). Entangling gates are presumed to require some kind of separate coupling element (purple). (c) A possible unit cell for dual-rail cavity qubits. Three beamsplitter couplers (purple) are required per unit cell, two of which are needed for entangling gates while the third actuates single qubit gates. The ancilla transmon (green) and readout resonator (grey) are only needed in half the unit cells, for the ancilla dual-rail qubits in a surface code.}
    \label{fig:unit cells}
\end{figure}

We first begin by accounting for the number of hardware elements needed to realize what we term the 'tile-able unit cell' for various possible superconducting qubit architectures as shown in Fig. \ref{fig:unit cells} and Tab. \ref{table:unit cell}. The tile-able unit cell includes not only the hardware needed to realize a physical qubit but also the couplings to surrounding qubits needed to realize a square lattice of physical qubits with nearest-neighbor connectivity, achieved solely by tiling the unit cell.

One way to compare the hardware resource overhead is to compare the number of transmon-like elements required to realize a surface code of a particular effective code distance. Compared to microwave resonators, transmons (which may be used as the physical qubit, tunable couplers or beamsplitter couplers as shown in Fig. \ref{fig:unit cells}) can be more resource intensive to fabricate at scale and so can serve as a point of reference for comparing different architectures. 

This is not the only way to compare different architectures. For instance, we could also compare the total number of readout lines or control lines instead. Nonetheless, using the total number of transmons as our metric, we can obtain useful insights into how dual-rail erasure qubits can lead to hardware savings despite having a more complicated unit cell.  

\begin{table*}[htb]
\label{table:unit cell}
\begin{tabular}{c|ccc}
\multirow{2}{*}{\textbf{\begin{tabular}[c]{@{}c@{}}Average number of elements \\ per tile-able unit cell\end{tabular}}} & \multicolumn{3}{c}{\textbf{Architecture}} \\ \cline{2-4} 
 & \multicolumn{1}{c|}{\begin{tabular}[c]{@{}c@{}}Transmons \\ w/ tunable couplers\\ (Frequency tunable)\end{tabular}} & \multicolumn{1}{c|}{\begin{tabular}[c]{@{}c@{}}Dual-rail transmons \\ w/ tunable couplers\cite{kubica2023erasure} \\ (Symmetric readout resonator)\end{tabular}} & \begin{tabular}[c]{@{}c@{}}Dual-rail cavities \\ w/ beamsplitter couplers \cite{teoh2023dual} \end{tabular} \\ \hline
 transmon-like elements & \multicolumn{1}{c|}{3} & \multicolumn{1}{c|}{5 (4)} & 3.5 \\
 readout lines & \multicolumn{1}{c|}{1} & \multicolumn{1}{c|}{3 (1)} & 0.5 \\
 2Q gate control lines & \multicolumn{1}{c|}{2} & \multicolumn{1}{c|}{2 (2)} & 2 \\
 single qubit control lines & \multicolumn{1}{c|}{1 (2)} & \multicolumn{1}{c|}{5 (3)} & 1 \\
 resonators & \multicolumn{1}{c|}{1} & \multicolumn{1}{c|}{3 (1)} & 2.5
\end{tabular}
\begin{caption}
    {Accounting for the number of hardware elements for three different possible superconducting qubit architectures to realize a square lattice of physical qubits with nearest-neighbor entangling gate connectivity. We attempt to do our unit cell accounting in a way which is as agnostic to the specific hardware implementation as possible. For the dual-rail cavity qubit architecture, only half the unit cells require an ancilla transmon and readout line, as reflected by values in the table.}
\end{caption}
\end{table*}

Now, let us fix the number of transmons and compare the effective distance $d_\mathrm{eff}$ when we use dual-rail transmon qubits with erasure noise~\cite{levine2023demonstrating,kubica2023erasure} and transmons as qubits with Pauli noise.
Suppose we have $N$ transmons, and we let $n$ be the number of transmons per tile-able unit-cell/physical qubit in a surface code. Physical qubits may include data and ancilla qubits, and $n$ includes tunable couplers as described in table~\ref{table:unit cell}. A surface code contains the order of $d^2$ physical qubits, thus, $N/n=d^2$. Furthermore, we know that for pure erasure noise, the effective distance is $d_\mathrm{eff}^{\mathrm{eras.}}=d$, while for Pauli noise, $d_\mathrm{eff}^{\mathrm{Pauli}}\approx d/2$. Thus, we quantify the gain factor in effective distance by using an erasure-based architecture with N transmons by the following calculation.
\begin{align}
    \frac{d_\mathrm{eff}^\mathrm{eras.}}{d_\mathrm{eff}^\mathrm{Pauli}}=\frac{\sqrt{N/n_\mathrm{eras.}}}{\frac{1}{2}\sqrt{N/{n_\mathrm{Pauli}}}}=2\sqrt{\frac{n_\mathrm{Pauli}}{n_\mathrm{eras.}}}
\end{align}
In a dual-rail transmon architecture, we assume that the number of transmons per physical qubit (including tunable couplers) is $n_\mathrm{eras}=5$. Using transmon qubits with tunable couplers requires $n_\mathrm{Pauli}=3$ transmons per physical qubit. Thus, our gain in effective distance by using dual-rail transmon qubits is $d_\mathrm{eff}^\mathrm{eras.}/d_\mathrm{eff}^\mathrm{Pauli}=2\sqrt{3/5}=1.55$. Therefore, for the same number of transmons $N$, there is only a $55\%$ increase in $d_\mathrm{eff}$ when using dual-rail transmon qubits. If we exclude the couplers in the transmon count, then we adjust the number of transmons per unit cell to $n_\mathrm{Pauli}=1$ and $n_\mathrm{eras}=3$. Then, the new $d_\mathrm{eff}$ ratio is $d_\mathrm{eff}^\mathrm{eras.}/d_\mathrm{eff}^\mathrm{Pauli}=2\sqrt{1/3}=1.15$. Thus, there is only a $15\%$ increase in $d_\mathrm{eff}$ when using dual-rail transmon qubits instead of transmon qubits with Pauli noise. This is in contrast to a $100\%$ increase in $d_\mathrm{eff}$, as naively expected by virtue of using erasure qubits instead of regular qubits with Pauli noise. This motivates the desire for hardware-efficient erasure architectures that can achieve a ratio closer to the theoretical gain in effective distance. Repeating this exercise for a dual-rail cavity architecture with built-in checks and assuming $n_\mathrm{eras.}=3.5$, we see that the gain in effective distance is $1.85$, or equivalently an $85\%$ increase in $d_\mathrm{eff}$.

Now suppose we want to achieve a target effective error distance $d_\mathrm{eff}$. We compare how many fewer transmons we need to reach this target distance on a surface code with dual-rail cavity qubits than transmon qubits. We set the effective distances equal to each other,
\begin{align}
    d_\mathrm{eff}^\mathrm{eras.}=d_\mathrm{eff}^\mathrm{Pauli}\\
    \sqrt{\frac{N_\mathrm{eras.}}{n_\mathrm{eras}}}=\frac{1}{2}\sqrt{\frac{N_\mathrm{Pauli}}{n_\mathrm{Pauli}}}\\
    \frac{N_\mathrm{Pauli}}{N_\mathrm{eras.}}=4 \left(\frac{n_\mathrm{Pauli}}{n_\mathrm{eras.}}\right)
\end{align}
Therefore, by using built-in erasure checks, we use $N_\mathrm{Pauli}/N_\mathrm{eras.}=4\left(3/3.5 \right)=3.4$ times fewer transmons. In other words, we need $29\%$ of the transmons to achieve the same $d_\mathrm{eff}$ compared to using transmon qubits with Pauli noise. If we were to use a dual-rail transmon architecture (with the physical qubit shown in \cite{levine2023demonstrating}) we would instead need 41\% of the transmons to achieve the same $d_\mathrm{eff}$.

\begin{figure*}[ht]
    \centering
    \includegraphics[width=\textwidth]{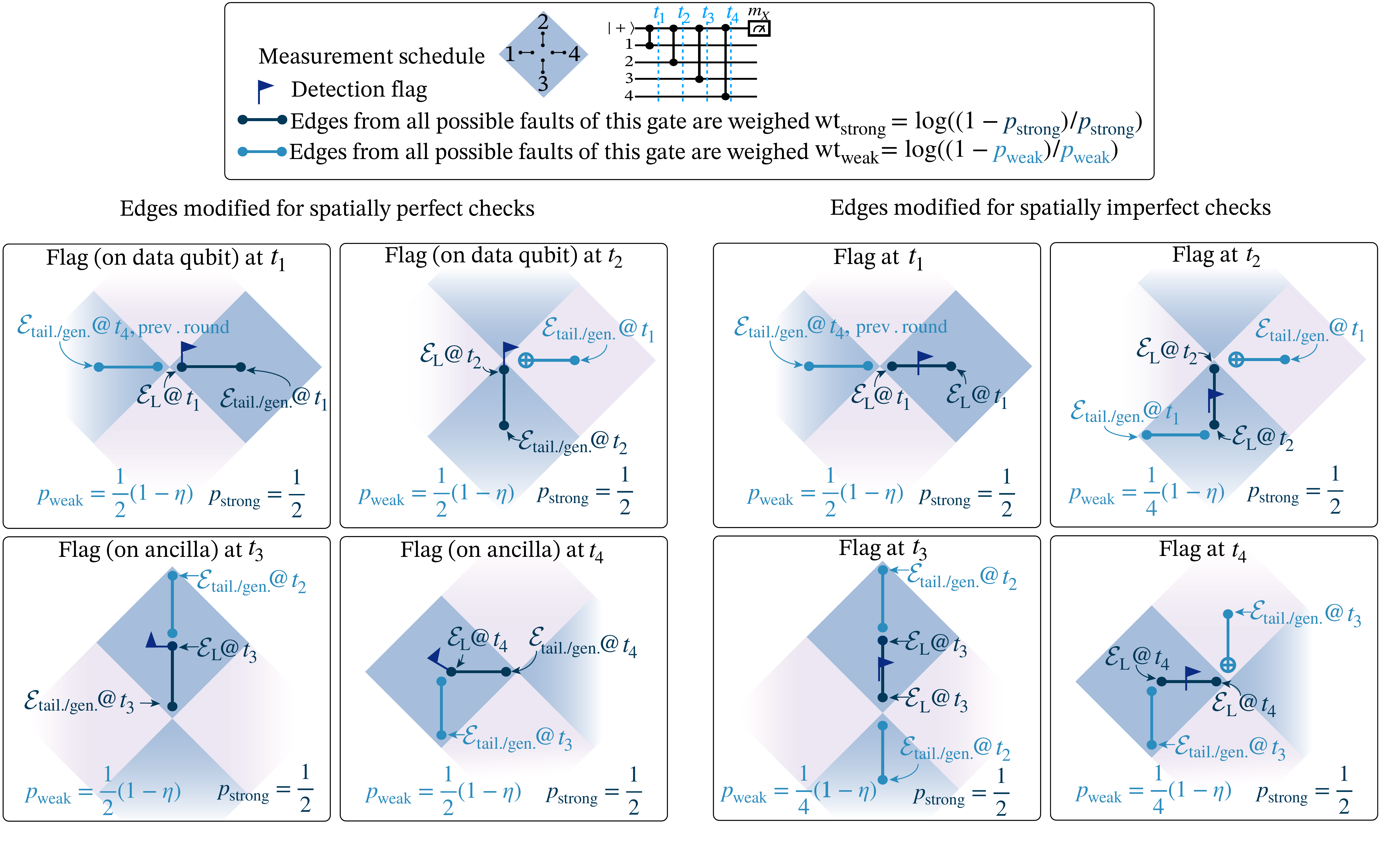}
    \caption{Given a flag from a spatially perfect, temporally imperfect check (left four boxes) or spatially imperfect, temporally imperfect check (right four boxes), we show which gates and their errors chosen from sets labeled $\mathcal{E}_{\mathrm{L}}$, $\mathcal{E}_{\mathrm{tail.}}$, or $\mathcal{E}_{\mathrm{gen.}}$ should be strongly erased (dark blue) or weakly erased (light blue). See Appendix~\ref{app: Decoding with imperfect erasures} for in-depth discussion.}
    \label{fig:strong_weak_eras}
\end{figure*}

\section{Decoding imperfect erasures}
\label{sec: decoding details}

In Section~\ref{sec: Noise and decoding when erasure detection is imperfect}, we described the population of errors depending on the spatial and temporal resolution of erasure checks. We will now describe in detail how these error mechanisms lead to edge weights that are defined on the decoding graph.
\subsection{Decoding preliminaries}

We decode $Z_\mathrm{L}$ and $X_\mathrm{L}$ information independently by defining two decoding graphs with nodes created from $X$-type and $Z$-type syndromes, respectively. In any decoding graph, an edge with a lower weight means that the error that creates syndromes at its ends is more probable. The standard way of weighing an edge is $\mathrm{wt}=\log(1-p)/p$ where $p$ is the probability of an error mechanism that creates syndromes at the ends of that edge~\cite{DennisTopoQuantumMemory}. The probability of Pauli errors due to noisy gates results in an edge weight distribution that is static in that it does not change with every memory experiment. If there are no Pauli errors, such as in the case where the only errors are erasures, then all edges have static weights $\mathrm{wt}=\infty$. Assuming this static, Pauli decoding graph as a starting point, we will talk about how the decoder uses the location of erasure flags to dynamically modify the weights of edges. 

\subsection{Perfect erasures}

In the case of perfect erasures, leakage is always accurately detected and immediately reinitialized into the computational mixed state $I/2$ following detection. This is modeled as a located depolarizing error. This depolarizing error results in specific syndromes which appear as two nodes in the X decoding graph with probability 1/2 and Z decoding graph with probability $1/2$. Thus, the edges between these syndromes are weighed $\mathrm{wt}_\mathrm{strong}=\log (1/2)/1/2=0$ and the syndromes are automatically paired up in a 0-parity cluster. We use the subscript strong to indicate that we know with certainty its syndromes should be joined within a 0-parity cluster in the decoding graph.

\subsection{Imperfect erasures}
\label{app: Decoding with imperfect erasures}
Now we discuss our decoder for imperfect erasure detection. This decoder takes the locations of flagged erasures and dynamically modifies edges depending on the leakage-induced Pauli model, spatial resolution, and temporal resolution $\eta$. In Figure~\ref{fig:strong_weak_eras}, given a spatial resolution (perfect or imperfect) and imperfect temporal resolution which is continuously varied by $\eta$, we illustrate which edges that require weight modification for a flag at each step of the stabilizer measurement circuit ($t_1-t_4$). The left four boxes indicate spatially perfect checks that flag individual qubits, while the right four boxes represent spatially imperfect checks that flag gates. For each flag in Figure~\ref{fig:strong_weak_eras}, we identify all gates with potential errors that occur at a time step immediately after the gate, at $t_1$, $t_2$, $t_3$, or $t_4$. These errors are chosen from either the sets $\mathcal{E}_\mathrm{L}$, $\mathcal{E}_\mathrm{gen.}$ or $\mathcal{E}_\mathrm{tail.}$ in accordance to how the models are defined in Section~\ref{sec: Noise and decoding when erasure detection is imperfect}. Note that due to separate $X$- and $Z$-syndrome decoding graphs, when collecting edges that result from faults drawn from $\mathcal{E}_{\mathrm{tail.}}$, edges may be asymmetrically modified in one graph compared to the other.

A dark blue \textit{strongly-erased} gate indicates our certainty that an error was definitely present after this gate, chosen out of the set of possible errors (denoted by $\mathcal{E}_{\mathrm{L}}$, $\mathcal{E}_{\mathrm{tail.}}$, or $\mathcal{E}_{\mathrm{gen.}}$). Conversely, a light blue \textit{weakly-erased} gate suggests that errors may have happened with a probability depending on $\eta$. Errors following a weakly erased gate are selected from either $\mathcal{E}_{\text{tail}}$ or $\mathcal{E}_{\text{gen}}$, depending on the leakage-induced Pauli model. Lower values of $\eta$ correspond to poorer temporal resolution, resulting in a higher likelihood of such errors after the weakly-erased gate. The probability that an error selected from these sets will lead to an edge on either the $X$- or $Z$-decoding graph is determined by the probabilities $p_\mathrm{strong}$ (for errors at strongly-erased gates) and $p_\mathrm{weak}$ (for errors at weakly-erased gates). Therefore, we assign weights to the edges on the $X$- and $Z$-decoding graphs that are formed by these error sets with either $\mathrm{wt_{strong}}=\log(1-p_{\mathrm{strong}})/p_{\mathrm{strong}}=0$ or $\mathrm{wt_{weak}}=\log(1-p_{\mathrm{weak}})/p_{\mathrm{weak}}>0$ for errors at strongly- or weakly- erased gates respectively. Figure~\ref{fig:strong_weak_eras} only display flags at CZ gates, but the same analysis would be applicable for any flagged CX gate by flipping the type (CX or CZ) of all weakly-erased gates. 

\begin{figure}[htb]
    \centering
    \includegraphics[width=\columnwidth]{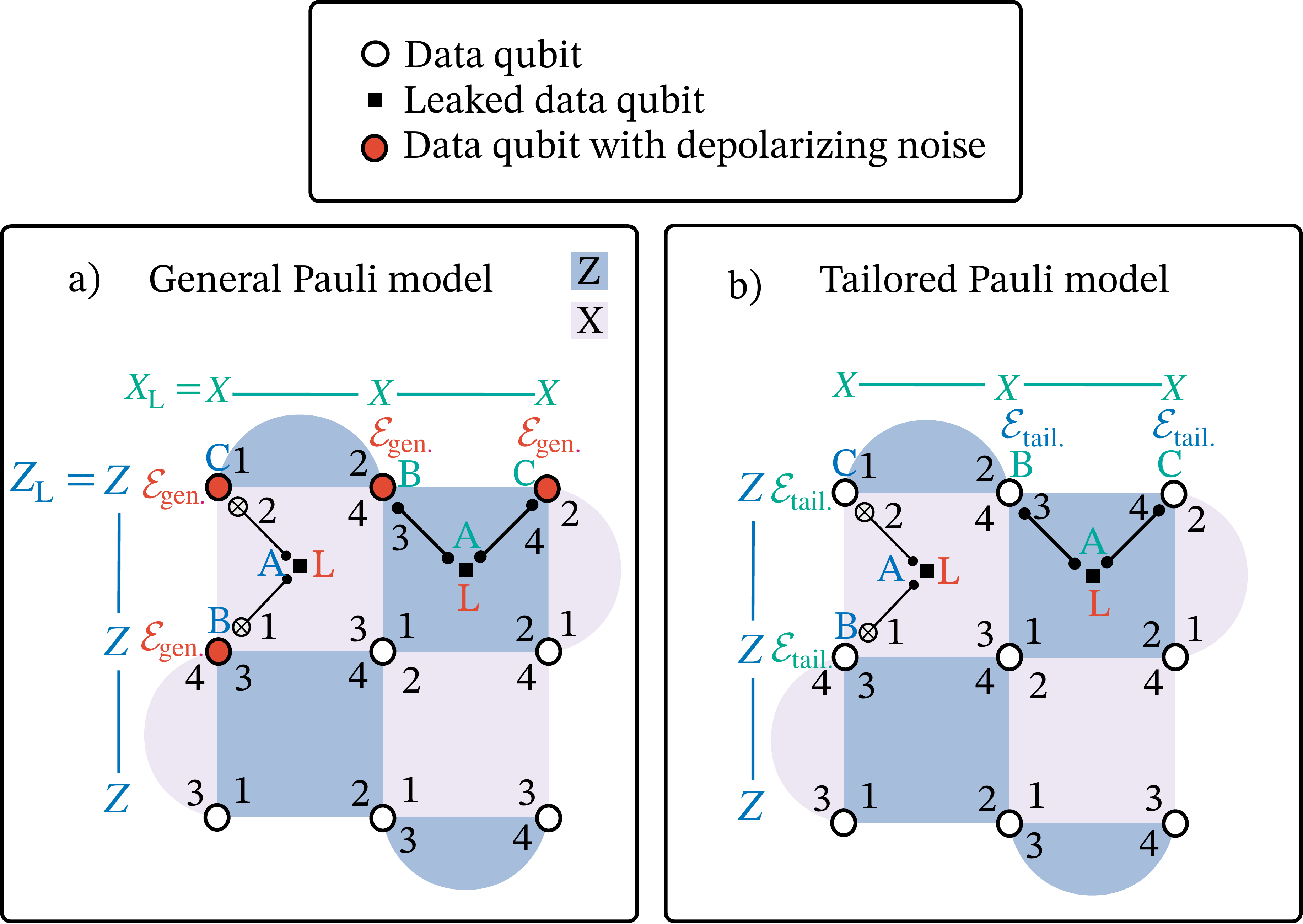}
    \caption{(a) General Pauli model. The effective error distance to a $Z_\mathrm{L}$ and $X_\mathrm{L}$ error should be reduced. Leakage of the ancilla at an $X$ stabilizer on the first CX gate leads to two depolarized qubits along the $Z_\mathrm{L}$ operator. Similarly, leakage of the ancilla at a $Z$ stabilizer third CZ gate leads to the same scenario along the $X_\mathrm{L}$ operator. (b) We show that under the tailored Pauli model, these cases do not lead to multiple errors with support along the logical operators. }
    \label{fig:rot_codedist_reduction}
\end{figure}

\section{Rotated codes}
\label{app:rotated codes}
Although the numerical analysis in this work is for the unrotated code, all the thresholds are also applicable to the rotated code. 
Furthermore, we now analytically examine the effect of imperfect checks on the effective error distance on rotated surface codes. As introduced in Section~\ref{sec: Noise and decoding when erasure detection is imperfect}, under delayed detection, a leaked qubit $A$ interacts with two qubits, $B$ and $C$, populating them with errors according to the general or tailored Pauli model. Moreover, we saw in Section~\ref{sec: effective error distance} that the effective distance is reduced to the Pauli noise case when qubits $B$ and $C$ have errors that lead to the same logical error. To determine if the rotated surface code experiences error distance reduction, in Figure~\ref{fig:rot_codedist_reduction}, we examine the standard measurement schedule~\cite{tomita2014lowdistance} of the rotated code, where numbers $1-4$ indicate the order of the two qubit gates. We see that the measurement schedule leads to configurations of qubits $B$ and $C$ along both $X_\mathrm{L}$ and $Z_\mathrm{L}$ operators. As a result, under the general Pauli model, qubits $B$ and $C$ are depolarized, thus reducing the effective distance to the $X_\mathrm{L}$ and $Z_\mathrm{L}$. Now let us examine what happens in the tailored Pauli model. Here, the errors on qubit $B$ are chosen from the set $\mathcal{E}_{\mathrm{tail.}}$, which will not have support on the same $X_\mathrm{L}$ or $Z_\mathrm{L}$ operator that qubit $C$ lies on. Thus, even in the rotated code, we expect the tailored Pauli model with delayed erasure checks to have unchanged effective error distance to both logical errors.

\begin{figure}[htb]
    \centering
    \includegraphics[width=\columnwidth]{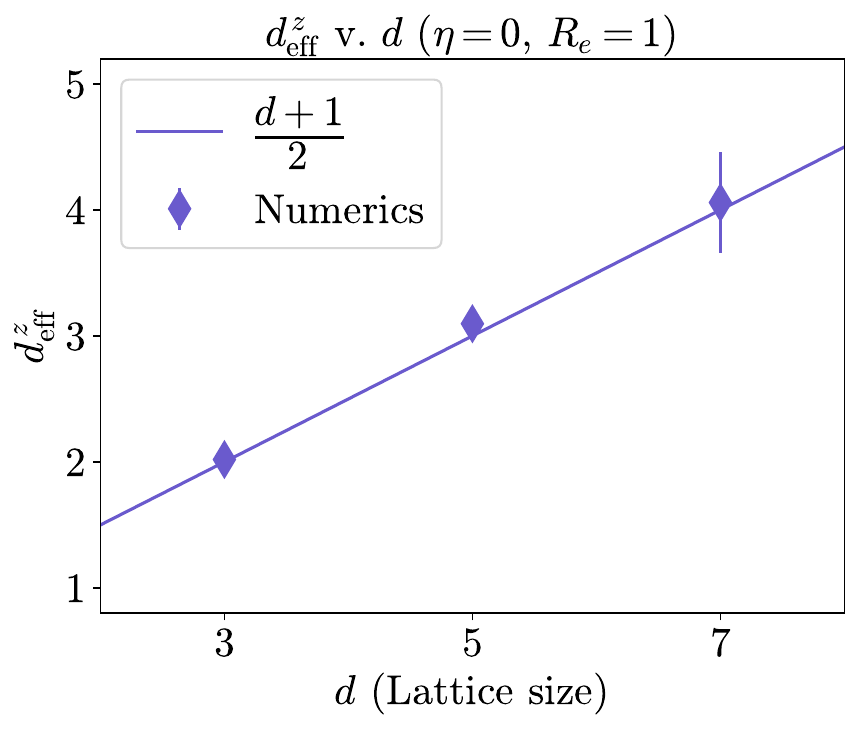}
    \caption{Numerically-obtained effective error distances to a $Z_\mathrm{L}$ error at $\eta=0$ for lattice sizes $d=3,5,7$. The solid line is the Pauli noise effective error distance, $(d+1)/2$.}
    \label{fig:codedist_vary_size}
\end{figure}

\section{Confirming effective distance scaling with different lattice sizes}
\label{app: lattice sizes}
In Figure~\ref{fig:codedist_vary_size}, we observe that the diminished effective distance to a $Z_\mathrm{L}$ error approaches the same error distance for Pauli noise, $(d+1)/2$ (solid line), for various unrotated lattice sizes $d=3$, $5$ and $7$ (diamonds), as predicted in Section~\ref{sec: noise and the effective distance}. Here we have considered the general Pauli model with spatially imperfect checks (see Figure~\ref{fig:unrot_codedist_reduction}~(a)) at temporal resolution $\eta=0$. As we have observed that the effective error distance is independent of spatial resolution (see Section~\ref{sec: effective error distance}), we do not expect these results to change for spatially perfect checks. The fact that the error distance approaches $(d+1)/2$ is unsurprising because each leakage event leads to potentially two data qubit $Z$ errors along the $Z_\mathrm{L}$. This means that it for an odd number of qubits in $Z_\mathrm{L}$, exactly $(d+1)/2$ leakage events may cause $Z$ errors on all the qubits in support of the logical. 

\section{3D threshold plots}
\label{app:3d threshold plots}

We present the thresholds in Figures~\ref{fig:phase_plots_general}~and~\ref{fig:phase_plots_tailored} in table-form. For constructing these figures, we evaluate the threshold at four different values of erasure fraction $\re=0.5, 0.7, 0.9, 1.0$ and five different values of $\eta=0.0, 0.5, 0.7, 0.9, 1.0$ each. Although not displayed, when there is only Pauli noise at $\re=0.0$, the threshold is $1.00 \pm 0.01 \%$ for all four tables/noise models.
\vfill\null
\begin{table}[h!]
\centering
\begin{tabular}{p{1.3cm}p{1.75cm} p{1.75cm} p{1.75cm} p{1.5cm}}
\\\toprule[1pt]
\multicolumn{5}{c}{General Pauli model \& Spatially Imperfect checks}                                                                                               \\ \toprule[1pt]
              & \textbf{$\re=0.52$} & \textbf{$\re=0.71$} & \textbf{$\re=0.91$} & $\re=1.0$ \\\toprule[1pt]
\textbf{$\eta=0.0$} & $1.40 \pm 0.01$   & $1.66 \pm 0.02$    & $2.11 \pm 0.02$    & $2.59 \pm 0.01$    \\ 
\multicolumn{1}{l}{\textbf{$\eta=0.5$}} & \multicolumn{1}{l}{$1.48 \pm 0.01$}    & \multicolumn{1}{l}{$1.80 \pm 0.02$}    & \multicolumn{1}{l}{$2.35 \pm 0.02$}    & $2.98 \pm 0.01 $   \\ 
\multicolumn{1}{l}{\textbf{$\eta=0.7$}} & \multicolumn{1}{l}{$1.53 \pm 0.01$}    & \multicolumn{1}{l}{$1.88 \pm 0.02$}    & \multicolumn{1}{l}{$2.50 \pm 0.02$}    & $3.27 \pm 0.01$    \\ 
\multicolumn{1}{l}{\textbf{$\eta=0.9$}} & \multicolumn{1}{l}{$1.57 \pm 0.02$}    & \multicolumn{1}{l}{$1.98 \pm 0.02$}    & \multicolumn{1}{l}{$2.83 \pm 0.02$}    & $3.93 \pm 0.01$    \\ 
\multicolumn{1}{l}{\textbf{$\eta=1.0$}} & \multicolumn{1}{l}{$1.60 \pm 0.02$}    & \multicolumn{1}{l}{$2.08 \pm 0.02$}    & \multicolumn{1}{l}{$3.12 \pm 0.04$}    & $5.09 \pm 0.02$    \\ \toprule[1pt]
\end{tabular}
\end{table}

\begin{table}[h!]
\centering
\begin{tabular}{p{1.3cm}p{1.75cm} p{1.75cm} p{1.75cm} p{1.5cm}}
\toprule[1pt]
\multicolumn{5}{c}{General Pauli model \& Spatially Perfect checks}                                                                                                         \\ \toprule[1pt]
       & \textbf{$\re=0.52$} & \textbf{$\re=0.71$} & \textbf{$\re=0.91$} & \textbf{$\re=1.0$} \\ 
\toprule[1pt]
\textbf{$\eta=0.0$} & $1.32 \pm 0.01$    & $1.63 \pm 0.02$    & $2.14 \pm 0.02$    & $2.96 \pm 0.02$    \\ 
\textbf{$\eta=0.5$} & $1.51 \pm 0.01$    & $1.86 \pm 0.02$    & $2.49 \pm 0.02$    & $3.33 \pm 0.02$    \\ 
\multicolumn{1}{l}{\textbf{$\eta=0.7$}} & \multicolumn{1}{l}{$1.54 \pm 0.01$}    & \multicolumn{1}{l}{$1.90 \pm 0.02$}    & \multicolumn{1}{l}{$2.66 \pm 0.02$}    & $3.57 \pm 0.03$   \\ 
\multicolumn{1}{l}{\textbf{$\eta=0.9$}} & \multicolumn{1}{l}{$1.59 \pm 0.02$}    & \multicolumn{1}{l}{$2.03 \pm 0.02$}    & \multicolumn{1}{l}{$2.88 \pm 0.02$}    & $4.17 \pm 0.03$    \\ 
\multicolumn{1}{l}{\textbf{$\eta=1.0$}} & \multicolumn{1}{l}{$1.60 \pm 0.02$}    & \multicolumn{1}{l}{$2.08 \pm 0.02$}    & \multicolumn{1}{l}{$3.12 \pm 0.04$}    & $5.09 \pm 0.02$    \\ \toprule[1pt]
\end{tabular}
\end{table}
\newpage

\begin{table}[ht]
\centering
\begin{tabular}{p{1.3cm}p{1.75cm} p{1.75cm} p{1.75cm} p{1.5cm}}
\toprule[1pt]
\multicolumn{5}{c}{Tailored Pauli model \& Spatially Imperfect checks}                                                                                                      \\ \toprule[1pt]
\multicolumn{1}{l}{}                 & \multicolumn{1}{l}{\textbf{$\re=0.52$}} & \textbf{$\re=0.71$} & \textbf{$\re=0.91$} & \textbf{$\re=1.0$} \\ \toprule[1pt]
\textbf{$\eta=0.0$} & $1.50 \pm 0.01$    & $1.85 \pm 0.02$    & $2.70 \pm  0.02$   & $3.49 \pm 0.02$    \\ 
\textbf{$\eta=0.5$} & $1.54 \pm 0.02$    & $1.95 \pm 0.02$    & $2.70 \pm 0.02$    & $3.79 \pm 0.02$    \\ 
\multicolumn{1}{l}{\textbf{$\eta=0.7$}} & \multicolumn{1}{l}{$1.55 \pm 0.02$}     & \multicolumn{1}{l}{$1.97 \pm 0.02$}    & \multicolumn{1}{l}{$2.80 \pm 0.02$}    & $4.02 \pm 0.02$    \\ 
\multicolumn{1}{l}{\textbf{$\eta=0.9$}} & \multicolumn{1}{l}{$1.55 \pm 0.02$}    & \multicolumn{1}{l}{$2.01 \pm 0.02$}    & \multicolumn{1}{l}{$2.96 \pm 0.02$}    & $4.58 \pm 0.02$    \\ 
\multicolumn{1}{l}{\textbf{$\eta=1.0$}} & \multicolumn{1}{l}{$1.60 \pm 0.02$}    & \multicolumn{1}{l}{$2.08 \pm 0.02$}    & \multicolumn{1}{l}{$3.12 \pm 0.04$}    & $5.09 \pm 0.02$    \\ \toprule[1pt]
\end{tabular}
\end{table}

\begin{table}[hb!]
\centering
\begin{tabular}{p{1.3cm}p{1.75cm} p{1.75cm} p{1.75cm} p{1.5cm}}
\toprule[1pt]
\multicolumn{5}{c}{Tailored Pauli model \& Spatially Perfect checks}
\\ \toprule[1pt]
                 & \textbf{$\re=0.52$} & \textbf{$\re=0.71$} & \textbf{$\re=0.91$} & \textbf{$\re=1.0$} \\ \toprule[1pt]
\textbf{$\eta=0.0$} & $1.48 \pm 0.01$    & $1.92 \pm 0.02$    & $2.84 \pm 0.02$    & $4.87 \pm 0.05$    \\ 
\multicolumn{1}{l}{\textbf{$\eta=0.5$}} & \multicolumn{1}{l}{$1.62 \pm 0.02$}    & \multicolumn{1}{l}{$2.16 \pm 0.02$}   & \multicolumn{1}{l}{$3.32 \pm 0.02$}    & $5.24 \pm 0.03$   \\ 
\multicolumn{1}{l}{\textbf{$\eta=0.7$}} & \multicolumn{1}{l}{$1.66 \pm 0.02$}   & \multicolumn{1}{l}{$2.18 \pm 0.02$}    & \multicolumn{1}{l}{$3.42 \pm 0.03$}    & $5.47 \pm 0.04$    \\ 
\multicolumn{1}{l}{\textbf{$\eta=0.9$}} & \multicolumn{1}{l}{$1.68 \pm 0.02$}    & \multicolumn{1}{l}{$2.30 \pm 0.02$}    & \multicolumn{1}{l}{$3.64 \pm 0.03$}    & $6.07 \pm 0.03$    \\ 
\multicolumn{1}{l}{\textbf{$\eta=1.0$}} & \multicolumn{1}{l}{$1.71 \pm 0.02$}    & \multicolumn{1}{l}{$2.33 \pm 0.03$}    & \multicolumn{1}{l}{$3.80 \pm 0.03$}    & $6.71 \pm 0.05$    \\ \toprule[1pt]
\end{tabular}
\end{table}
\vfill\null
\newpage~\newpage

\end{document}